\documentstyle[pra,aps,multicol,epsfig]{revtex}

\input epsf.tex

\newcommand{\be}{\begin{equation}}
\newcommand{\ee}{\end{equation}}
\newcommand{\ba}{\begin{eqnarray}}
\newcommand{\ea}{\end{eqnarray}}
\newcommand{\ban}{\begin{eqnarray*}}
\newcommand{\ean}{\end{eqnarray*}}

\newcommand{\braket}[2]{\mbox{$ \langle #1 | #2 \rangle $}}
\newcommand{\sandwich}[3]{\mbox{$ \langle #1 | #2 | #3 \rangle $}}
\newcommand{\ket}[1]{\mbox{$ | #1 \rangle $}}
\newcommand{\bra}[1]{\mbox{$ \langle #1 | $}}

\newcommand{\si}{\sigma}
\newcommand{\demi}{\frac{1}{2}}

\newcommand{\myone}{\leavevmode\hbox{\small1\normalsize\kern-.33em1}}

\newcommand{\moy}[1]{\langle #1 \rangle}

\begin{document}

\title{Zero-Error Attacks and Detection Statistics in the Coherent One-Way Protocol for Quantum Cryptography}
\author{Cyril Branciard$^{1}$, Nicolas Gisin$^{1}$, Norbert L\"utkenhaus$^{2}$, Valerio Scarani$^{1}$}
\address{$^1$ Group of Applied Physics, University of Geneva; 20, rue de
l'Ecole-de-M\'edecine, CH-1211 Geneva 4, Switzerland
\\ $^2$ Institute for Quantum Computing, University of Waterloo; 200 University Ave.
W., Waterloo, ON Canada N2L 3G1}
\date{\today}
\maketitle

\begin{abstract}

This is a study of the security of the Coherent One-Way (COW)
protocol for quantum cryptography, proposed recently as a simple
and fast experimental scheme. In the zero-error regime, the
eavesdropper Eve can only take advantage of the losses in the
transmission. We consider new attacks, based on unambiguous state
discrimination, which perform better than the basic beam-splitting
attack, but which can be detected by a careful analysis of the
detection statistics. These results stress the importance of
testing several statistical parameters in order to achieve higher
rates of secret bits.

\end{abstract}

\begin{multicols}{2}

\section{Introduction}

First proposed by Bennett and Brassard in 1984 (BB84 protocol,
\cite{bb84}), quantum cryptography has attracted a lot of attention,
as means of realizing a useful task (key distribution for secret
communication) based on the superposition principle of quantum
physics. One of the features, that makes quantum cryptography
appealing, is the possibility of implementing it with present-day
technology. After several years devoted to more and more elaborated
realizations of the BB84 protocol \cite{review}, people gained in
confidence, and started devising new protocols that are tailored for
practical implementations. A new class of such protocols are {\em
distributed phase reference schemes} \cite{dps,COW_1,COW_2}, where
the signals have overall phase-relationships to each other which is
expected to protect against some loss-related attacks, such as the
photon-number splitting attack, in a similar way as the strong phase
reference in the original Bennett 1992 (B92) protocol \cite{b92}
does. These new protocols are providing new challenges for
theorists, as we can no longer identify individual signals, and so
the usual security proof techniques do not apply. It is important to
understand how we prove the security, and the context of the present
work is to show limitations of secure rates by showing specific
attacks that can be performed by an eavesdropper.

In a protocol like BB84, each bit is coded in a qubit: Alice
prepares a photon in a given state which codes (say) for 0 and
sends it to Bob; then, she prepares another photon in another
state which codes (say) for 1, and sends it, and so on. In short,
each quantum signal codes for one bit. For this kind of protocols,
powerful security proofs have been derived for the case where the
quantum signal is a single photon
\cite{sp,rennerthesis,krausrenner} or a weak coherent pulse
\cite{inamori,gllp}. But one can also code a bit in the {\em
relative phase} between any two successive coherent pulses: in
such a protocol (called {\em differential phase shift}) the first
bit is in the phase between pulse one and pulse two, the second
bit in the phase between pulse two and pulse three, and so on
\cite{dps}. Thus, each pulse participates to the coding of two
bits and is coherent with all the other pulses: {\em there is a
unique quantum signal, the string of all the pulses, which codes
for the whole string of bits}.

The search for security bounds for such schemes is an important
research activity in theoretical quantum cryptography. In this
paper, we study a protocol of the same kind called {\em Coherent
One-Way (COW)} \cite{COW_1,COW_2}, which will be explained in detail
later. We present new attacks on this protocol based on unambiguous
state discrimination. These attacks take advantage of the fact that,
on the one hand, the coding of COW makes use of empty pulses and, on
the other hand, coherence is checked only between successive pulses:
in particular then, no coherence is checked between all that comes
before and all that comes after an empty pulse. Therefore, if Eve
can be sure that a given pulse was empty, she can make an attack
that breaks no observed coherence. The attacks that we have found do
not introduce any errors in the statistical parameters that are
usually checked, the quantum bit error rate (QBER) and the
visibility of an interferometer; but they do introduce modifications
in other statistical parameters, which Alice and Bob could check as
well. The main message of this paper is that the COW protocol should
include additional statistical checks. Of course, since we describe
specific attacks, in this paper we derive only {\em upper bounds}
for security (i.e., more powerful attacks may exist).

The paper is organized as follows. In Section \ref{seccow} we recall
the definition of the COW protocol and introduce our working
assumptions. Section \ref{sec3} presents unambiguous state
discrimination (USD) strategies on three and four successive pulses,
and the detection rates for the COW protocol that Bob would observe
if Eve applied those strategies. In Section \ref{secrates}, we
present our main results: an attack that combines three USD
strategies and that preserves all the observed detection rates in
Bob's detectors. Section \ref{concl} is a conclusion. In the
Appendices, we provide the security study for a three-state protocol
that is the analog of the COW protocol if the coherence between bits
would be broken (Appendix \ref{App_3_state}) and for the
beam-splitting attack considered as a collective attack (Appendix
\ref{App_BS_Holevo}); we also present the detailed calculations for
the best attack that we have found (Appendix \ref{appqs}) and an
attack that becomes possible if Alice and Bob would make a too
limited statistical analysis (Appendix \ref{appexample}); finally,
we suggest a feasible modification of the COW protocol that would
improve its security (Appendix \ref{App_empty_decoy}).

\section{The COW protocol}
\label{seccow}

\subsection{The protocol}

\begin{center}
\begin{figure}
\includegraphics[width=7cm]{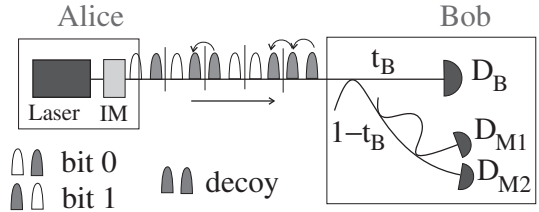}
\vspace{2mm} \caption{Schematic description of the COW protocol. A
continuous, phase-stabilized coherent laser beam is sent through
an intensity modulator (IM) that shapes discrete pulses, while
preserving the coherence. See text for all other details.}
\label{figcow}
\end{figure}
\end{center}

The idea of the COW protocol is to have a very simple {\em data
line} in which the raw key is created, protected by the
observation of quantum interferences in a {\em monitoring line}.
We review here its features, referring to Refs \cite{COW_1,COW_2}
for a more comprehensive discussion of motivations and practical
issues. The protocol is schematized in Fig.~\ref{figcow}.

Alice produces a train of equally spaced coherent pulses. The {\em
logical bit 0} is encoded in the sequence
$\ket{0}_{2k}\ket{\alpha}_{2k-1}$ of a non-empty pulse at time
$t_{2k-1}$ followed by an empty one at time $t_{2k}$; the {\em
logical bit 1} in the opposite sequence
$\ket{\alpha}_{2k}\ket{0}_{2k-1}$. We write $\mu = |\alpha|^2$ the
mean photon number in a non-empty pulse. Alice produces each bit
value with probability $\frac{1-f}{2}$; with probability $f$, she
sends out the {\em decoy sequence}
$\ket{d}=\ket{\alpha}_{2k}\ket{\alpha}_{2k-1}$, which does not
encode any bit value. The coherence time of Alice's laser is very
large, so that the quantum signal cannot be divided bitwise, because
there is phase coherence between any two non-empty pulses. In other
words, there is a single quantum signal, defined by Alice's list,
e.g. \ba \ket{"...0d01..."}&=&
\ket{...:0\alpha:\alpha\alpha:0\alpha:\alpha 0:...} \ea (from now
on, the colon represents the bit separation). The coherence across
different bits is crucial to this scheme --- a protocol that uses
the same coding of bits, but in which there is no distributed
coherence, is presented in Appendix \ref{App_3_state}.

Alice and Bob are connected by a quantum channel of length $\ell$,
whose transmission coefficient is $t=10^{-\alpha_{att} \ell /
10}$; the parameters $\alpha_{att}$, whose units are dB/km, is
called attenuation coefficient.

Bob's detection is completely passive. At the entrance of Bob's
device, an asymmetric coupler sends a fraction $t_B$ of the
photons into the data line, and the remaining fraction $1-t_B$
into the monitoring line. The data line consists of a single
photon counter $D_B$: the logical bits 0 and 1 are discriminated
by measuring the time of arrival (this gives indeed the best
unambiguous state discrimination between the states
$\ket{0}\ket{\alpha}$ and $\ket{\alpha}\ket{0}$). The errors on
the data line give the quantum bit error rate (QBER, $Q$). The
monitoring line contains a stabilized unbalanced interferometer
and two photon counters $D_{M1}$, $D_{M2}$. In the interferometer,
the delayed half of each pulse is recombined by the non-delayed
half of the next pulse: if the two pulses were non-empty, the
interference is arranged in such a way that $D_{M2}$ should never
click. The cases where two successive pulses are non empty are (i)
the decoy sequences, in which case the coherence is within the bit
separation, and (ii) a logical bit 1 followed by a logical bit 0,
in which case the coherence is across the bit separation. In each
of these cases separately ($s=d$ or $s=1-0$), Alice and Bob can
estimate the errors through the visibility
$V_{s}=\frac{p(D_{M1}|s)-p(D_{M2}|s)}{p(D_{M1}|s)+p(D_{M2}|s)}$
where $p(D|s)$ is the probability that detector $D$ has fired at a
time corresponding to a $s$ sequence.

For the estimation of the visibilities and of the counting
statistics, Bob announces (i) in which two-pulse sequence he had a
detection in the data line, and (ii) at which times he had a
detection in $D_{M1}$ and $D_{M2}$. Alice tells Bob which items of
the data line must be discarded because they correspond to decoy
sequences; on her side, she estimates $V_d$ and $V_{10}$ and the
counting statistics. Finally, $Q$ is estimated as usual by Bob
revealing some of the bits of the data line.

The amount of information gathered by Eve is estimated through
$Q$, $V_{d}$, $V_{10}$, but not only: the monitoring of other
statistical quantities may provide much better estimates.
Specifically, it is important to monitor {\em detection rates}, as
we show in this paper. Finer checks could involve the monitoring
of the frequency of each bit value and of many-bit strings, the
rate at which any two or all three detectors fire, etc.

\subsection{Detection statistics in the zero-error case}

In this work, we consider only attacks that introduce no errors in
the state parameters of the coding ($Q=0$, $V=1$), and that can
therefore be detected only by looking at the statistics of the
photon counters. Among the statistical parameters, we focus on
{\em detection rates}. We suppose that all three Bob's detectors
have the same quantum efficiency $\eta$ and no dark counts. We
also work in the {\em trusted-device scenario}, i.e. the
inefficiency of the detector is not given to Eve. Under these
assumptions, the expected detection rates are the following:
\begin{itemize}
\item In detector $D_B$, one can estimate the detections due to
"bits" and those due to "decoy sequences" (detection rate per two
time-slots): \ba D_{B,bit}^t &=& (1-f)
(1 - e^{-\mu \,t\, t_B \eta})\,,\label{dbbit}\\
D_{B, decoy}^t &=& 2\,f\, (1 - e^{- \mu\, t \,t_B \eta})\,;
\label{dbdecoy}\ea of course, the total detection rate in this
detector is \ba D_{B}^t&=&D_{B,bit}^t+D_{B,decoy}^t\,.
\label{db}\ea

\item  In detectors $D_{M1}$ and $D_{M2}$, one can estimate two
different detection rates. (i) The detection rates at time
$t_{2k}$ correspond to interference between two pulses within a
bit sequence. The logical bits produce random outcomes, while the
decoy sequences interfere constructively in $D_{M1}$ (recall
$V=1$): \ba D_{M1, 2k}^t &=& (1-f)D_{rand} + f D_{int}\,,\\D_{M2,
2k}^t &=& (1-f) D_{rand} \ea where $D_{rand}=1 - e^{-\mu t (1-t_B)
\eta / 4}$ and $D_{int}=1 - e^{- \mu t (1-t_B) \eta}$. (ii) The
detection rates at time $t_{2k+1}$ correspond to interference
between two pulses across the bit separation. Constructive
interference appears in $D_{M1}$ in the cases $1-0$, $1-d$, $d-0$
and $d-d$, i.e. with probability $(1+f)^2/4$; in the case $0-1$
there is no photon, so no detection, in the other cases the
outcome is random:
\ba D_{M1, 2k+1}^t &=& \frac{1-f^2}{2}\, D_{rand} \,+\, \frac{(1+f)^2}{4}\, D_{int}\,, \\
D_{M2, 2k+1}^t &=& \frac{1-f^2}{2}\,D_{rand}\,.\ea
\end{itemize}
Now, since $t_B$ has been calibrated by Bob, these six detection
rates depend only on two parameters, namely $f$ and $x\equiv
e^{-\mu t\eta}$. Bob can verify that the observed detection rates
are consistent in themselves, and with the expected values of $f$
and $x$.

About other statistical quantities that can be checked by Alice
and Bob: in the attacks that we consider below, the coincidence
rates are not really a concern, the bit values are equally
probable; but the many-bit statistics are somehow biased and may
reveal the attacks.

\subsection{Zero-error attacks}

In the ideal situation that we consider (zero-error, i.e. $Q=0$,
$V=1$), the eavesdropper can take advantage only of the losses in
the channel, whose transmission is $t$. Here we characterize the
full set of attacks that Eve can have performed, if Alice and Bob
observe zero errors.

The simplest attack is {\em beam-splitting (BS) attack}: Eve
simulates the lossy channel by extracting the $(1-t)$ fraction of
the signal with a beam-splitter, and sends the expected fraction
$t$ to Bob on a lossless line. Since a beam-splitter is strictly
equivalent to losses, this attack is always possible and is
impossible to detect by monitoring the data of Alice and Bob.
Thus, this attack sets an obvious upper bound on the achievable
secret key rate. We analyze it in detail in Appendix
\ref{App_BS_Holevo}, improving over the study of
Ref.~\cite{COW_2}. Though it is unavoidable, the BS attack is not
very powerful: it would be a very good point for a protocol, if it
could be shown that this attack is the only possible one in the
absence of errors.

The BS attack is an example of {\em attacks that preserve the
mode}, while possibly changing the statistics of the photon
numbers; these attacks always belong to the class of zero-error
attacks. In distributed phase reference schemes, each photon
belongs to an extended mode that encodes the coherence.
Specifically, in the case of differential phase shift, the mode is
$A^{\dagger}=\frac{1}{\sqrt{N}}
\sum_{j=1}^N\,e^{i\varphi_j}\,a_{j}^{\dagger}$ where
$a_{j}^{\dagger}$ creates a photon in the $j$-th pulse
\cite{waks}. In the case of COW, the extended mode is \ba
A^{\dagger}&\propto&\sum_{k=1}^N\,a_{k,s_k}^{\dagger} \ea where
$s_k\in\{0,1,d\}$ defines the nature of the $k$-th two-pulse
sequence, and the creation operators are
$a_{k,0}^{\dagger}=a_{2k-1}^{\dagger}$,
$a_{k,1}^{\dagger}=a_{2k}^{\dagger}$ and
$a_{k,d}^{\dagger}=a_{2k-1}^{\dagger}+a_{2k}^{\dagger}$.

The attacks that preserve the extended mode would be the only
zero-error attacks if Alice and Bob would check all the coherence
relations. In the case of COW however, Alice and Bob check the
coherence only on two {\em successive} pulses: in particular, no
coherence is checked between all that comes before and all that
comes after an empty pulse. Therefore, if Eve can be sure that a
given pulse was empty, she can make an attack that breaks the
coherence at the location of that pulse. More generally, Eve can
try and distinguish a sequence of $n$ pulses that begins and ends
with an empty pulse: if she succeeds, she can then resend photons
belonging to this $n$-slots mode ("partial mode"). All these
attacks must use {\em unambiguous state discrimination (USD)}. In
this paper we study examples of such attacks.

The list of zero-error attacks is now complete. To see it, we note
that any photon received from Bob is either one of the photons
originally sent by Alice (which then belongs to the original
extended mode), or a new photon created by Eve (in which case she
must have known exactly in which partial mode to send it). In
particular, the photon-number splitting (PNS) attack \cite{pns} is
never a zero-error attack for the schemes under study
\cite{COW_2,dpspns}: since any two non-empty pulses are coherent,
any attempt of measuring the number of photons on a finite number
of pulses breaks some coherence and contributes to errors.

\section{Unambiguous State Discrimination on Three and Four Pulses}
\label{sec3}

\subsection{Generalities}
\label{ssgen}

The attacks that we study are based on {\em unambiguous state
discrimination (USD)}. Suppose the set of possible states is known
(cryptography is a natural example \cite{dusek}): the unambiguous
discrimination of any state $\ket{\psi}$ in the set is possible if
and only if this state is linearly independent from all the other
states in the set \cite{chefles}. For the present study, we just
need to identify {\em one} state $\ket{\psi}$ in the set; therefore,
we consider measurements with only {\em two} outcomes: the
unambiguous identification and the inconclusive
outcome~\cite{noteUSD}. In this case, the optimal USD strategy is as
follows: in the subspace formed by the states of the set, one
selects $\ket{\phi}$ as the state orthogonal to all but
$\ket{\psi}$, and performs the von Neumann measurement
$\big\{P_{c}=\ket{\phi}\bra{\phi}, P_{\perp}=\myone-P_c\big\}$. If
the state was not $\ket{\psi}$, the result is certainly $\perp$; so
if the result is $c$, the state was certainly $\ket{\psi}$. Given
that the state is $\ket{\psi}$, the conclusive result $c$ happens
with probability $p_c=\left|\braket{\psi}{\phi}\right|^2$.

Specifically, Eve wants to discriminate a given finite sequence of
pulses from all the other possible ones; the chosen sequence must
be such that the first pulse and the last one are empty. When the
result is conclusive, she can prepare and forward the same
sequence to Bob; when the result is inconclusive, we suppose that
she blocks everything (finer strategies are possible, but we
neglect them \cite{notefine}). By definition, such an attack
leaves $Q=0$ and $V=1$, because Bob receives something only when
Eve is sure to forward the same sequence as Alice sent, and
because no observable coherence has been broken thanks to the
empty pulses \cite{notedps}. However, Eve introduces losses,
because the conclusive result is only probabilistic; and,
according to the state she actually discriminates and forwards,
Bob's statistics are also modified.

Our goal in what follows is to quantify the amount of information
that Eve obtains and to analyze how Bob's statistics are affected,
for some examples of USD attacks on the COW protocol.
Specifically, we are going to present three USD attacks. These
three attacks can be alternated with one another without
introducing errors. Eve can also avoid errors by stopping the USD
attacks after a successful discrimination. However, she cannot
avoid the risk of errors if she resumes the attack again. What she
can do, is to attack large blocks, then to stop also for a large
block, then resume and so on: this way, the events in which Eve
risks introducing an error have almost zero statistical weight (in
particular, they can be overwhelmed by dark counts and other
imperfections, which are neglected here).

\subsection{USD3: Attack on Three Pulses}

The USD3 attack is defined as follows: Eve takes three pulses that
come from Alice and wants to discriminate unambiguously the
sequence $\ket{0 \alpha 0}$ from the other possible three-pulses
sequences. When the discrimination is successful, she forwards
some photons (not necessarily a coherent state) in the central
time-slot; when the result is not conclusive, she doesn't forward
anything. One can see immediately that this attack doesn't
introduce any errors in the data line, preserves the randomness of
the bit value, and doesn't make detector $D_{M2}$ of Bob's
monitoring line click when it shouldn't. The limitation of this
attack is that Eve never forwards anything when Alice had sent two
successive non empty pulses; so, if this attacks is performed
systematically, Alice and Bob notice that no decoy sequences have
been detected, nor do they have any data to estimate $V$.

\subsubsection{Discriminating $\ket{0 \alpha 0}$}

Eve wants to discriminate the state $\ket{0 \alpha 0}$ from the
other possible states, which are the following: \be \ket{0 0
\alpha}\,,\; \ket{0 \alpha \alpha}\,,\; \ket{\alpha 0 0}\,,\;
\ket{\alpha 0 \alpha}\,,\; \ket{\alpha \alpha 0}\,,\; \ket{\alpha
\alpha \alpha}\,. \label{6states}\ee Note that the sequence $\ket{0
0 0}$ is never sent by Alice. Moreover, the sequences $\ket{0 0
\alpha}$ and \ket{\alpha 0 0} can be sent only if the bit separation
is between the two empty pulses; given that Eve knows the position
of the separation, she therefore has only to discriminate between
$\ket{0 \alpha 0}$ and five other states.

For each case, the six possible states are linearly independent. As
a consequence, there is a state in this 6-dimensional subspace which
is orthogonal to the five other possible states: this state is (in
both cases) \ba \ket{\psi_{0 \alpha 0}} &=&
\frac{1}{1-\chi^2}(\ket{0 \alpha 0} - \chi \ket{0 \alpha \alpha} -
\chi \ket{\alpha \alpha 0} + \chi^2 \ket{\alpha \alpha \alpha})
\label{psi3}\ea where $\chi = \braket{0}{\alpha} = e^{-|\alpha|^2/2}
= e^{-\mu/2}$. Eve performs a projective measurement which separates
$\ket{\psi_{0 \alpha 0}}$ from the subspace orthogonal to it.
Conditioned on the fact that the state $\ket{0 \alpha 0}$ was sent
by Alice, the probability of a conclusive result is $\left|\braket{0
\alpha 0}{\psi_{0 \alpha 0}}\right|^2 = (1 - \chi^2)^2=(1 -
e^{-\mu})^2$.

\subsubsection{Detection rates in COW for USD3}

Let us compute the detection rates in Bob's detectors when Eve
performs the USD3 attack. Eve forwards something to Bob with
probability \ba p_{concl}^{0 \alpha 0} &=&
\left(\frac{1-f}{2}\right)^2 (1 - e^{-\mu})^2\,. \label{pok3}\ea
We denote by $\Pi(p)=1-\moy{(1-p)^n}_{{\cal E}}$ the average
detection probability of the state $\ket{{\cal E}}$ that Eve
forwards, as a function of the single-photon probability detection
$p$. In particular, $\Pi(p)=p$ if Eve forwards a single photon,
$\Pi(p)\approx 1$ if she forwards a bright pulse. The detection
rates on the detector $D_B$ are \ba D_{B,bit}^{(3)}
&=& \frac{2}{3}\, p_{concl}^{0 \alpha 0}\,\Pi\left(t_B\eta\right) \,,\label{db3}\\
D_{B, decoy}^{(3)} &=& 0\,. \ea The factor $\frac{2}{3}$ comes
from the fact that we compute the detection rate per bit, i.e. for
two time slots, while the attack was performed on three pulses.
The detection rates on the monitoring line are just random clicks,
since two successive pulses are never sent, and so we find \ba
D_{M1,
2k}^{(3)} &=& D_{M2, 2k}^{(3)} = D_{M1, 2k+1}^{(3)} = D_{M2, 2k+1}^{(3)}\nonumber \\
&=& \frac{2}{3} p_{concl}^{0 \alpha 0} \Pi\left((1-t_B)
\frac{\eta}{4}\right) \; , \ea where the factor $\frac{1}{4}$ in
the transmission probability comes from the fact that each photon
has the "choice" between two paths in the interferometer, and the
"choice" between two detectors.

\subsection{USD4a: A First Attack on Four Pulses}

The USD4a attack is defined as follows: Eve takes four pulses
coming from Alice that correspond to two bits, and she wants to
discriminate the sequence $\ket{0 \alpha : \alpha 0}$ from the
other possible sequences. As before, when Eve successfully could
discriminate this sequence, she forwards photons in the two middle
time slots, making sure they will interfere correctly in Bob's
monitoring line, while when she couldn't discriminate this
sequence she doesn't forward anything.

Again, this attack doesn't introduce any bit error, and doesn't
make the detector $D_{M2}$ click when it shouldn't. Contrary to
USD3, $V$ can be estimated, but only from $1-0$ bit sequences: no
decoy sequences are ever forwarded.

\subsubsection{Discriminating $\ket{0 \alpha : \alpha 0}$}

Eve wants to discriminate the sequence $\ket{0 \alpha : \alpha 0}$
from the other possible following states that Alice could send:
\be \begin{array}{c} \ket{0 \alpha : 0 \alpha}, \ket{0 \alpha :
\alpha
\alpha}, \ket{\alpha 0 : 0 \alpha}, \ket{\alpha 0 : \alpha 0}, \\
\ket{\alpha 0 : \alpha \alpha}, \ket{\alpha \alpha : 0 \alpha},
\ket{\alpha \alpha : \alpha 0}, \ket{\alpha \alpha : \alpha
\alpha}. \label{8states} \end{array} \ee In the subspace defined
by the nine possible states, the state which is orthogonal to the
eight states listed in (\ref{8states}) is \be \begin{array}{rcl}
\ket{\psi_{0 \alpha : \alpha 0}} & = & \frac{1}{1-\chi^2}(\ket{0
\alpha : \alpha 0} - \chi \ket{0 \alpha : \alpha \alpha} \\ & &
\qquad - \chi \ket{\alpha \alpha : \alpha 0} + \chi^2 \ket{\alpha
\alpha : \alpha \alpha})\,.
\end{array} \ee Eve performs a projective measurement which separates
$\ket{\psi_{0 \alpha: \alpha 0}}$ from the subspace orthogonal to
it. Conditioned on the fact that the state $\ket{0 \alpha: \alpha
0}$ was sent by Alice, the probability of a conclusive result is
$\left|\braket{0 \alpha : \alpha 0}{\psi_{0 \alpha : \alpha
0}}\right|^2 = (1 - \chi^2)^2$. This is the same probability as
obtained before, in the discrimination of three-pulse state
$\ket{0\alpha 0}$.

\subsubsection{Detection rates in COW for USD4a}

Let us compute the detection rates in Bob's detectors when Eve
performs the USD4a attack. Eve forwards something to Bob with
probability $p_{concl}^{0 \alpha : \alpha 0}$ which, as we just
stressed, is given by (\ref{pok3}). The detection rates on the
detector $D_B$ are \ba D_{B,bit}^{(4a)} &=& \frac{1}{2}\,
p_{concl}^{0 \alpha:\alpha 0}\,\Pi\left(t_B\eta\right) \,,\label{db4a}\\
D_{B, decoy}^{(4a)} &=& 0\,.\ea The factor $\frac{1}{2}$ comes
from the fact that we compute the detection rate per bit, i.e. for
two time slots, while the attack was performed on 4 pulses. We
have also assumed that Bob's detectors have no dead time
\cite{notedead}.

The detection rates on the monitoring lines behave differently,
according to the time. The detections at times $t_{2k}$ are just
random, since there are no decoy sequences and consequently no
interference between pulses within a bit sequence: \ba D_{M1,
2k}^{(4a)} &=& D_{M2, 2k}^{(4a)} = \frac{1}{2} \,p_{concl}^{0
\alpha : \alpha 0}\,\Pi\left((1-t_B) \frac{\eta}{4}\right)\,. \ea
On the contrary, when Eve forwards something, there is always a
coherence across the bit separation; therefore the detections at
times $t_{2k+1}$ exhibit full interference effects: \ba D_{M1,
2k+1}^{(4a)} & = & \frac{1}{2}\, p_{concl}^{0 \alpha : \alpha 0}
\,\Pi\left((1-t_B)
\frac{\eta}{2}\right) \\
D_{M2, 2k+1}^{(4a)} & = & 0\,. \ea

\subsection{USD4b: A Second Attack on Four Pulses}

The two attacks USD3 and USD4a share the same feature, namely,
that no decoy sequences ever reach Bob. In order to pass as much
unnoticed as possible, Eve could be obliged to alternate those
attacks with another one, in which decoy sequences are sent. We
consider the simplest one, in which Eve wants to discriminate
$\ket{0 : \alpha \alpha : 0}$ from the other possible sequences.
Again, the colon represents the bit separation: contrary to USD4a,
now the four pulses are across three bit sequences.

One realizes immediately that this is a curious attack: if
performed systematically, Eve would forward only decoy sequences,
so no raw key would be created! As we said, it is interesting to
consider it only as a part of a more complex attack, in which Eve
would alternate it with the attacks we have already presented.

\subsubsection{Discriminating $\ket{0 : \alpha \alpha : 0}$}

One might expect that the probability of conclusive result is the
same as before. But this is not the case: there are now more
possible sequences, across the 3 bits, that Alice could send.
Specifically, Eve wants to discriminate the sequence $\ket{0 :
\alpha \alpha : 0}$ from the following eleven states: \be
\begin{array}{c} \ket{0 : 0 \alpha : 0}, \ket{0 : \alpha 0 : 0},
\\ \ket{0 : 0 \alpha : \alpha}, \ket{0 : \alpha 0 : \alpha}, \ket{0 : \alpha \alpha : \alpha},
\\ \ket{\alpha : 0 \alpha : 0}, \ket{\alpha : \alpha 0 : 0}, \ket{\alpha : \alpha \alpha : 0},
\\ \ket{\alpha : 0 \alpha : \alpha}, \ket{\alpha : \alpha 0 : \alpha}, \ket{\alpha : \alpha \alpha : \alpha}.
\label{11states} \end{array} \ee The state orthogonal to these
eleven states is \ba \ket{\psi_{0 : \alpha \alpha : 0}} & =
&\frac{(1+\chi^2)\phi(\alpha\alpha)-\chi\left[
\phi(0\alpha)+\phi(\alpha 0)\right]}{\sqrt{1-\chi^4}} \ea where we
have written \ba\phi(X)&=&\frac{\ket{0 X 0} - \chi \ket{0 X
\alpha}- \chi \ket{\alpha X 0} + \chi^2 \ket{\alpha X
\alpha}}{1-\chi^2}\,. \label{def_phi} \ea Conditioned on the fact
that the state $\ket{0 :\alpha \alpha: 0}$ was sent by Alice, the
probability of a conclusive result is $\left|\braket{0 : \alpha
\alpha : 0}{\psi_{0 : \alpha \alpha : 0}}\right|^2 = \frac{(1 -
\chi^2)^3}{1 + \chi^2}$. Note that this is much smaller than the
value $(1 - \chi^2)^2$ obtained in the previous examples:
specifically, for $\mu\ll 1$, it goes as $\demi \mu^3$ (three
photons) instead of $\mu^2$ (two photons).

\subsubsection{Detection rates in COW for USD4b}

Eve forwards something to Bob with probability \ba p_{concl}^{0:
\alpha \alpha: 0}&=& f \,\left(\frac{1-f}{2}\right)^2 \,\frac{(1 -
e^{-\mu})^3}{1 + e^{-\mu}}\,.\ea The detection rates on the
detector $D_B$ are \ba D_{B,bit}^{(4b)} &=& 0 \,,\\
D_{B, decoy}^{(4b)} &=& \frac{1}{2}\, p_{concl}^{0 :\alpha \alpha:
0}\,\Pi\left(t_B\eta\right)\ea with the same factor $\demi$ as
discussed for the USD4a attack. Detections in the monitoring line
behave just the opposite way as they did for the USD4a attack: \ba
D_{M1, 2k}^{(4b)} & = & \frac{1}{2}\, p_{concl}^{0 :\alpha \alpha:
0}
\,\Pi\left((1-t_B) \frac{\eta}{2}\right)\,, \\
D_{M2, 2k}^{(4b)} & = & 0\,;\\ D_{M1, 2k+1}^{(4b)} &=& D_{M2,
2k+1}^{(4b)} = \frac{1}{2} \,p_{concl}^{0 :\alpha \alpha:
0}\,\Pi\left((1-t_B) \frac{\eta}{4}\right)\,. \ea In summary,
there is an obvious symmetry between the USD4a and USD4b attacks.
However, the fact that $p_{concl}^{0: \alpha \alpha:
0}<p_{concl}^{0 \alpha: \alpha 0}$ introduces an important
difference. In fact, the need for sending some decoy sequences is
very costly for Eve: she has to perform sometimes a very
inefficient attack, which moreover gives her no information on the
key (she knows that the decoy sequence was preceded by a bit 1 and
followed by a bit 0, but she does not send anything to Bob apart
from the decoy sequence itself, so these two bits cannot be
detected).

\section{Combining the three USD attacks}
\label{secrates}

In the previous Section, we have described an attack where Eve
forwards "bits" (USD3), an attack where she forwards "coherence
across the bit separation" (USD4a), and an attack which forwards
"decoy sequences" (USD4b). These are zero-error attacks as far as
the state parameters are concerned ($Q=0$, $V=1$), but each one
taken separately introduces deviations from the expected detection
rates. Here we show that, provided $f\lesssim 0.236$, Eve can
alternate among the three attacks in order to simulate all the
expected detection rates.

\subsection{Definition of the attack}
\label{sec:attack_def} The attack that we consider (with no claim
of optimality) is constructed as follows. Eve performs USD3 with
probability $q_1$, USD4a with probability $q_2$, and USD4b with
probability $q_3$. With probability $q_0$, she just forwards the
pulses through a lossless channel ($t = 1$). Recall that Eve can
alternate as she likes among the USD attacks, but she must not
stop and resume them too often (see end of paragraph \ref{ssgen}).

We suppose that this is all she does, so that \ba q_0 + q_1 + q_2
+ q_3 &=& 1\,.\label{norm}\ea We want all detection rates to be
the expected ones: the six rates $D = D_{B,bit}$, $D_{B,decoy}$,
$D_{M1,2k}$, $D_{M2,2k}$, $D_{M1,2k+1}$ or $D_{M2,2k+1}$ must be
such that \ba q_0 D^{t=1} + q_1 D^{(3)} + q_2 D^{(4a)} + q_3
D^{(4b)}&=& D^t \,.\label{requirement} \ea We make two further
assumptions, namely (i) that Eve forwards always a single photon
when she has got a conclusive result \cite{notecoinc}, in
particular then $\Pi(p)=p$; and (ii) that we can work in the limit
$\mu\eta\ll 1$, so that we can linearize all the detection rates
$D^t$. In this case, an analytical solution can be found (Appendix
\ref{appqs}), that reads \ba
q_0&=&\frac{\mu t F-1}{\mu F-1}\label{solqs}\\
q_j&=&\frac{\mu(1-t)F_j}{\mu F-1}\quad(j=1,2,3) \ea where \ba
F_1&=&\frac{3(1-4f-f^2)}{4p_{concl}^{0\alpha
0}}\,=\,\frac{3(1-4f-f^2)}{(1-f)^2}\frac{1}{(1-e^{-\mu})^2}\,,\\
F_2&=&\frac{(1+f)^2}{p_{concl}^{0\alpha:\alpha
0}}\,=\,4\left(\frac{1+f}{1-f}\right)^2\frac{1}{(1-e^{-\mu})^2}\,,\\
F_3&=& \frac{4f}{p_{concl}^{0:\alpha\alpha:
0}}\,=\,\frac{16}{(1-f)^2}\frac{1+e^{-\mu}}{(1-e^{-\mu})^3}\,,
\label{f3}\\
F&\equiv&F_1+F_2+F_3\,=\,\frac{1}{(1-f)^2} \frac{32-{\cal
F}(1-e^{-\mu})}{(1-e^{-\mu})^3}\label{sumf}\ea with ${\cal
F}=9+4f-f^2$. Note that, while $F_{2}$ and $F_3$ are always
strictly positive, for $F_1$ to be non-negative one must have
$f\leq \sqrt{5}-2\approx 0.236$: this means that Eve cannot
reproduce the detection rates with this attack if a large fraction
of decoy sequences is used.

\subsection{Upper Bound on the Secret Key Rate}

We can now compute the secret key rate that can be extracted by
Alice and Bob in the presence of the attack just described. We
consider the case of one-way classical post-processing, and use
the Csisz\`ar-K\"orner formula \cite{ck} \ba R&=&
D_{B,bit}^t\,\left[I(A:B)-\min\left(I(A:E),
I(B:E)\right)\right]\ea where $H$ is Shannon entropy, $I(X:Y)$ is
mutual information, and by definition of our attacks we have \ba
D^t_{B,bit}&=& q_0 D_{B,bit}^{t=1} + q_1 D_{B,bit}^{(3)} + q_2
D_{B,bit}^{(4a)}\,.\ea The use of the Csisz\`ar-K\"orner formula
can be justified by an argument analog to the one used in
Ref.~\cite{moroder}: the USD attack immediately gives a
decomposition of the data into those on which Eve has full
information (i.e. those on which the USD attack has been applied
and has given conclusive result) and those on which Eve has no
information at all (i.e. those that have been sent over the ideal
channel). In this case, the Csisz\`ar-K\"orner formula gives a
tight bound if Alice and Bob were sure that Eve is performing
exactly that attack; since this is not proved (there might be
better attacks compatible with the observed statistics), the value
of $R$ that we compute is an {\em upper bound on the secret key
rate that can be extracted with one-way post-processing}.

\begin{center}
\begin{figure}
\includegraphics[width=8cm]{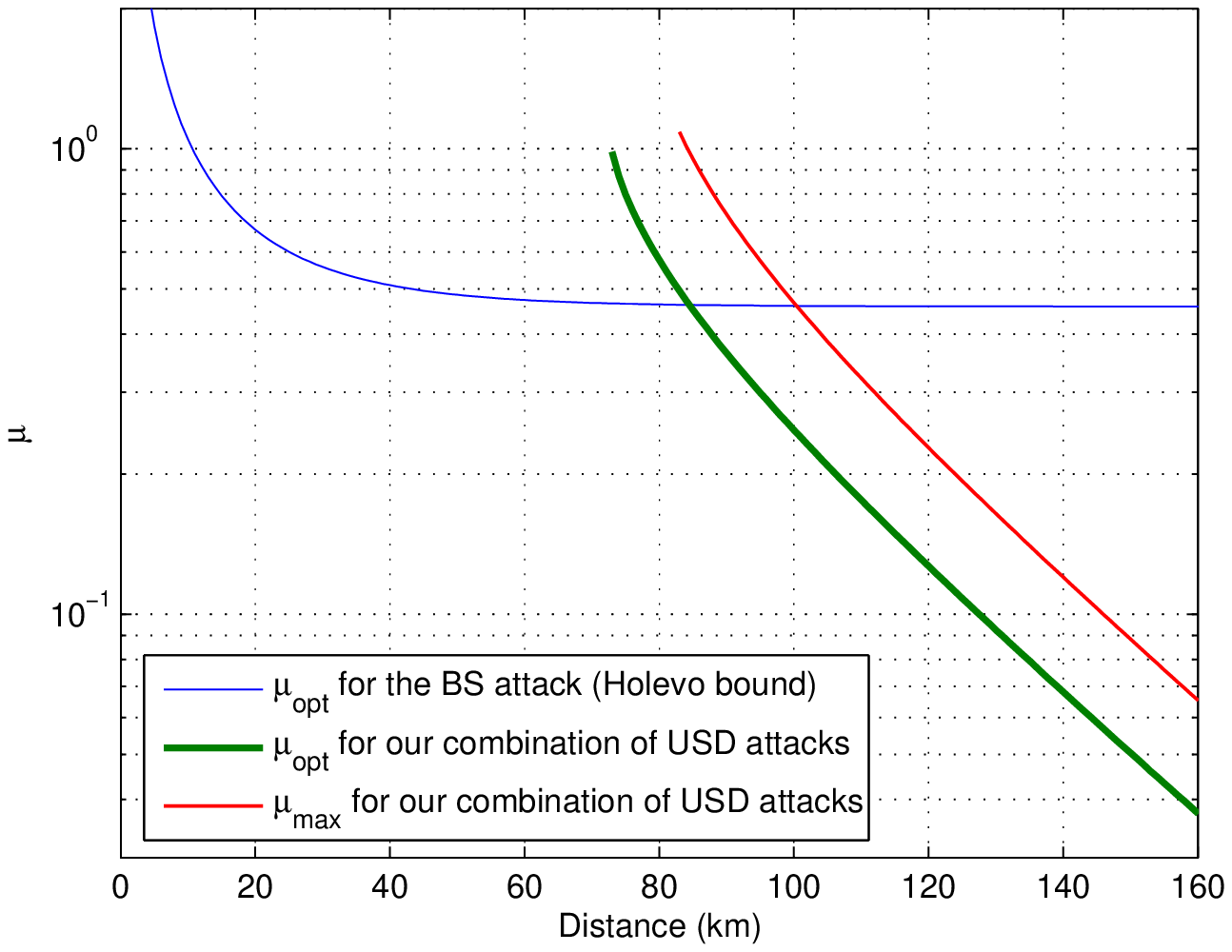} \vspace{2mm}
\includegraphics[width=8cm]{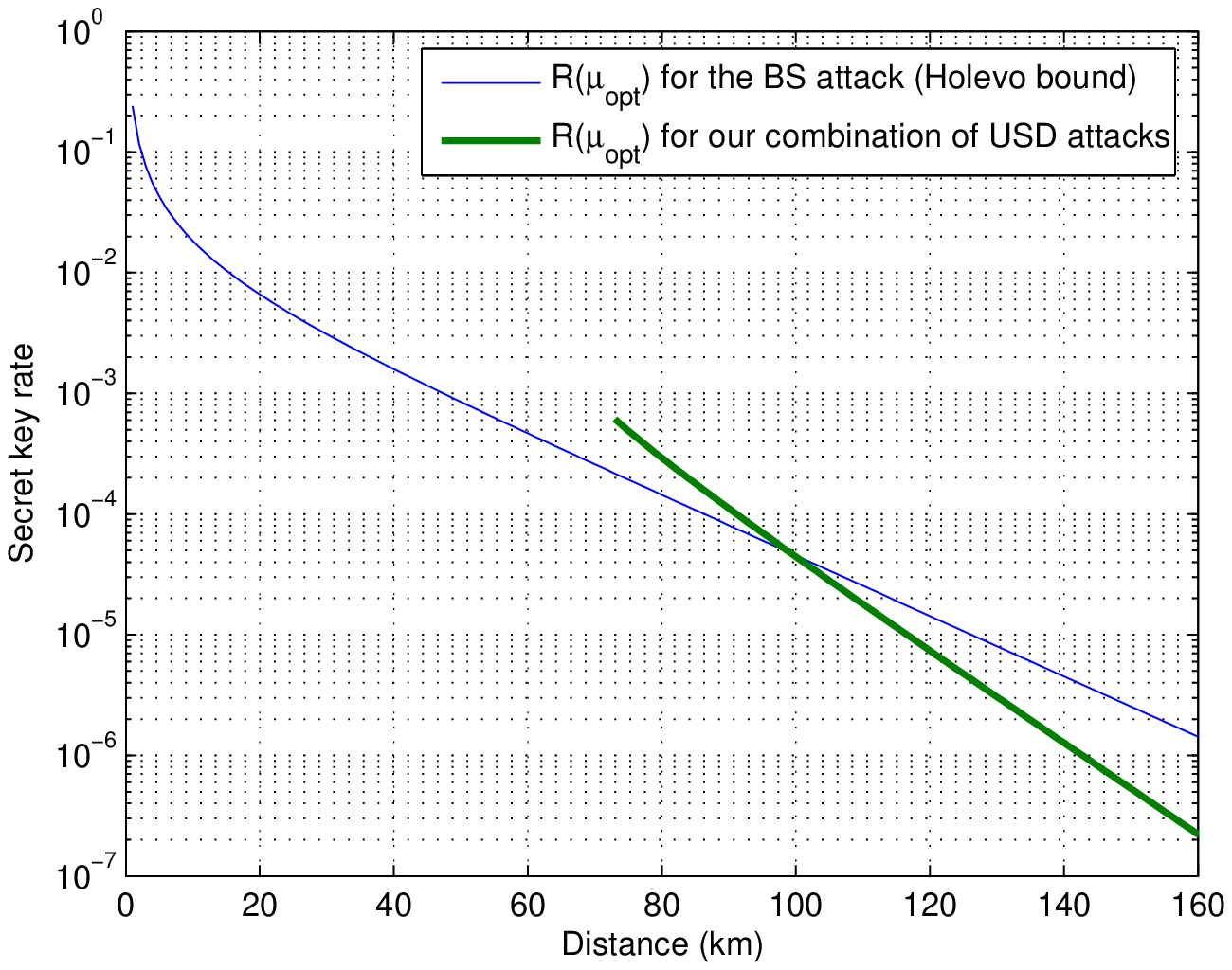} \caption{USD
attack that reproduces the detection rates: optimal mean photon
number $\mu_{opt}$ (upper graph) and corresponding secret key rate
$R$ (lower graph) as a function of the Alice-Bob distance $\ell$.
The attack is compared to the Holevo bound on the beam-splitting
attack. Parameters: $\eta = 0.1$, $\alpha_{att} = 0.25$ dB/km, $f =
0.1$, $t_B \simeq 1$.} \label{figmain}
\end{figure}
\end{center}

Now, on the one hand, since there are no errors in the state,
whenever Bob detects something in $D_B$ (other than a decoy
sequence) he learns correctly Alice's bit: \ba I(A:B)&=&1\,.\ea This
implies $I(A:E)=I(B:E)$. On the other hand, Eve has full information
on the bits that she attacked and forwarded and were detected in
$D_B$, and she has no information in all the other cases: \ba
I(A:E)&=& \frac{q_1 D_{B,bit}^{(3)} + q_2
D_{B,bit}^{(4a)}}{D^t_{B,bit}}\,.\ea This gives the expected
results, namely that Alice and Bob have secrecy if and only if the
bit was not attacked by Eve: \ba R(\mu) &=& q_0
D_{B,bit}^{t=1}\,=\,\frac{\mu t F(\mu)-1}{\mu
F(\mu)-1}\,\mu\,t_B\eta(1-f)\,. \ea As usual, Alice and Bob choose
the value of $\mu$ that maximizes $R$. Another meaningful parameter
is $\mu_{max}$, the critical value such that $R=0$ (that is,
$q_0=0$: Eve can perform her attack on all the bits). The
calculation of $\mu_{opt}$, $R(\mu_{opt})$ and $\mu_{max}$ has been
done numerically; the results are shown in Fig.~\ref{figmain}. These
parameters can also be estimated analytically in the limit $\mu\ll
1$, using $F(\mu) \approx \frac{1}{(1-f)^2} \frac{32}{\mu^3}\,$ and
therefore $q_0 \approx t - \frac{(1-f)^2}{32}\,\mu^2$; it yields \ba
\mu_{opt}& \approx &
\frac{4\sqrt{6}}{3(1-f)} \,\sqrt{t}\,,\label{muopt}\\
R(\mu_{opt})& \approx & \frac{8\sqrt{6}}{9} \,t_B\eta\,t^{3/2}\,,
\label{Rmax}\\\mu_{max}& \approx &\sqrt{3}\,\mu_{opt}\,.\ea For long
distances, these analytical estimation are in close agreement with
the numerical optimization.

In Fig.~\ref{figmain}, our attack is compared to the Holevo bound
on the beam-splitting (BS) attack computed in Appendix
\ref{App_BS_Holevo}. As we can see in the lower graph, the BS
attack is more powerful than ours for $\ell\lesssim 100$km; by
referring to the upper graph, we note a discontinuity in
$\mu_{opt}$. This is due to the fact that we have not considered a
mixture between our attack and the BS attack; if we had considered
it, the transition between the two would have been smooth.

\subsection{Comments on the result}

We have described a specific attack, which introduces no errors in
the state parameters, and which reproduces all the expected
detection rates as well. Let's comment on the results.

To the attack, as we have studied it, many limitations can be
found. First, this attack is not a real concern as of today: in
fact, it outperforms the BS attack only for $\ell\gtrsim 100$km
(Fig.~\ref{figmain}), which is anyway the typical limiting
distance when dark counts are taken into account \cite{COW_2}.
Second, the attack is not entirely undetectable with the actual
setup: even though all the detection rates are reproduced, one
could check other statistical parameters, which would behave in an
unexpected way. For instance, since decoy sequences are always
forwarded in the form $\ket{0:\alpha\alpha:0}$, Alice and Bob can
realize that the two pulses before a decoy sequence that they
detect always encodes a logical bit 1, and the two pulses after
the decoy sequence always encodes a logical bit 0. Finally, as
seen in Sec. \ref{sec:attack_def}, Alice and Bob could simply
choose $f>0.236$, and the attack that we studied becomes
impossible.

A further interesting point is that the power of the attack can be
further reduced by a hardware modification, which keeps the
simplicity of the experimental realization: it simply amounts at
adding {\em empty decoy sequences}. The idea is that, by adding a
new kind of signal, the conclusive probability of USD become
smaller, because Eve has to distinguish the desired state among a
larger set. The analysis is done in Appendix
\ref{App_empty_decoy}; the intuition is confirmed: by adding empty
decoy sequences, we obtain a decrease $R(\mu_{opt})\propto
t^{4/3}$ [Eq.~(\ref{rempty})] at long distances, which is slower
than $R(\mu_{opt})\propto t^{3/2}$ given in Eq.~(\ref{Rmax}). Note
that other hardware modifications would help as well, in
particular adding interferometers that monitor coherence across
more than one pulse; but these would make the experiment more
complicated \cite{noteinterf}.

All these arguments can be made as an objection to the importance
of our attack. However, that precise attack is only an example:
there is no claim of optimality. There is some room for
improvement even on USD strategies with three and four pulses
\cite{notefine}, and we have not studied USD attacks on more than
four pulses. Another concern is that we don't have any estimate of
the robustness of our result when the precision of the statistical
estimates of Alice and Bob decreases. Here, we have worked without
dark counts and in the limit of an infinite sequence: the presence
of dark counts and the finite-size effects, obviously present in
any real experiment, may blur the statistics. Eve's attack may
become much more serious if she is asked to guarantee only an
approximation of the expected detection rates, or to reconstruct
only a smaller set of statistical quantities. A simple example of
what can happen if Alice and Bob do not make a careful enough
statistical estimate is given in Appendix \ref{appexample}.

\section{Conclusion}
\label{concl}

In conclusion, we have studied the security of the COW protocol in
the regime of zero error in the state parameters ($Q=0$, $V=1$).
In this regime, Eve can take advantage only of the losses; while
the beam-splitting attack is always possible, because it preserves
the collective mode in which all photons have been encoded, we
addressed the existence of more powerful attacks.

We have indeed found examples of other zero-error attacks, which
however introduce some modifications in the statistics observed by
Bob. We have presented an attack that preserves all the detection
rates and can be detected only by looking at correlations between
two or more bits. This attack becomes relevant only for large
distances ($\ell\gtrsim 100$ km for typical values).

These results show that, both in the experiment and in the
theoretical search of lower bounds for security, higher secret key
rates can be achieved if the COW protocol includes several tests
of Bob's statistics. We conjecture that the beam-splitting attack
is the only possible one in the zero-error limit provided Alice
and Bob analyze {\em all} statistics of their data.

\section*{Acknowledgements}

C.B. acknowledges hospitality in IQC Waterloo, where part of this
work was realized. We acknowledge financial support from the
European Project SECOQC and from the Swiss NCCR "Quantum
Photonics".

\appendix

\section{Three-state protocol}
\label{App_3_state}

Here we describe a three-state protocol, that was inspired by the
study of the COW protocol. If the coherence across the bit
separations in COW would be broken, the protocol could be seen as
a implementation with weak coherent pulses of a standard
three-state protocol for qubits. The qubits states thus obtained
are \ba
\begin{array}{lcl} \ket{+z} &\equiv& \ket{0}\\ \ket{-z}
&\equiv& \ket{1}\\ \ket{+x} &\equiv& \frac{1}{\sqrt{2}}(\ket{0} +
\ket{1})\,.
\end{array} \ea Each state of the $Z$
basis is sent with probability $(1-f)/2$; it codes a bit value,
and the errors in these measurements give the quantum bit error
rate (QBER) $Q$. The third state, belonging to the $X$ basis, is
sent with probability $f$; it allows to estimate a visibility $V$.

In this appendix we give a quick overview of security studies for
this protocol, relying mainly on Ref.~\cite{krausrenner}, to which
we refer for the justification of the methods. An independent
study of this three-state protocol has been realized recently by
Fung and Lo with different techniques \cite{funglo}.

\subsection{Single photon case}

\subsubsection{Quick review of the approach}

In Ref.~\cite{krausrenner}, a lower bound on the secret-key rate
for a general class of quantum key distribution protocols using
one-way classical post-processing has been derived. Remarkably,
the bound can be computed considering only two-qubit density
operators $\sigma_{AB}$ \cite{random_transfo}: \ba r &\geq&
\inf_{\sigma_{AB} \in \Gamma_{Q,V}} S(A|E) - H(A|B)\nonumber\\ &=&
\inf_{\sigma_{AB} \in \Gamma_{Q,V}} 1 - S(\sigma_{AB})
\label{lower_bound}\ea where $S$ is the Von Neumann entropy, $H$
is the Shannon entropy, and the second line is obtained when Eve
holds a purification of $\si_{AB}$ which is a usual assumption in
quantum cryptography. The set $\Gamma_{Q,V}$ is the set of
two-qubit Bell-diagonal density operators which are compatible
with the measured QBER $Q$ and visibility $V$. Our goal is to
characterize this set, and then to perform the minimization in
Eq.~(\ref{lower_bound}). This is done by using the
entanglement-based description of the three-state protocol, and
considering the most general attack that Eve can perform on a
qubit that goes from Alice to Bob.

\subsubsection{Qubit pairs shared by Alice and Bob}

Let us first consider the equivalent entanglement-based version of
the three-state protocol: Alice prepares the state \be
\ket{\Psi_{AB}} = \sqrt{1-f} \ket{\Phi^+}_{AB} + \sqrt{f} \ket{D}_A
\ket{+x}_B \ee where we used the standard notation $\ket{\Phi^+} =
\frac{1}{\sqrt{2}}(\ket{00} + \ket{11})$, and where $\ket{D}_A$ is a
state orthogonal to $\ket{0}_A$ and $\ket{1}_A$ (Alice's system is
therefore 3-dimensional); she keeps the first system and sends the
second one to Bob.

On her system, Alice performs a projective measurement in order to
prepare Bob's state. When Alice gets the result $\ket{0}_A$ (which
she does with probability $\frac{1-f}{2}$), she prepares the state
$\ket{0}_B$ for Bob; when she gets $\ket{1}_A$ (with probability
$\frac{1-f}{2}$), she prepares the state $\ket{1}_B$ for Bob;
finally, when she gets $\ket{D}_A$ (with probability $f$), she
prepares a decoy sequence $\ket{+x}_B$ for Bob.

The system $B$ that goes from Alice to Bob through the quantum
channel can be attacked by Eve. Let us describe her action by a
super operator ${\cal E} = \{E_j\}$. The  state shared by Alice
and Bob after the transmission of system $B$ is then
\ba \rho_{AB} & = & \mathcal{E}(\ket{\Psi_{AB}}\bra{\Psi_{AB}})\nonumber \\
& = & \sum_{j} \myone_A \otimes E_j \ket{\Psi_{AB}} \bra{\Psi_{AB}}
\myone_A \otimes E_j^\dag\,. \ea After the public communication,
Alice and Bob know which  systems led to bits of the key (when Alice
obtained either $\ket{0}_A$ or $\ket{1}_A$ and Bob measured in the
$Z$ basis), and which systems came from decoy sequences (when Alice
obtained $\ket{D}_A$ and Bob measured in the $X$ basis). They have 2
sets of systems in the states : \ba \rho_{AB}^{bit} & = &
(\ket{0}\bra{0} + \ket{1}\bra{1})_A
 \ \rho_{AB} \ (\ket{0}\bra{0} + \ket{1}\bra{1})_A\nonumber
 \\
& = & (1-f) \sum_{j} \myone_A \otimes E_j \ket{\Phi_{AB}^+}
\bra{\Phi_{AB}^+} \myone_A \otimes E_j^\dag \\
\rho_{AB}^{decoy} & = & \ket{D}\bra{D}_A \ \rho_{AB} \
\ket{D}\bra{D}_A \nonumber\\
& = & f \sum_{j} \myone_A \otimes E_j \ \ket{D,+x}_{AB} \bra{D,+x}
\ \myone_A \otimes E_j^\dag\,. \ea We shall write $\rho =
\widetilde{\rho}_{AB}^{bit} =\frac{1}{1-f}{\rho}_{AB}^{bit}$ and $
\widetilde{\rho}_{AB}^{decoy} =\frac{1}{f} \,{\rho}_{AB}^{decoy}$
the corresponding normalized states. Note that $\sqrt{2} \ket{D}
\bra{+x}_A \otimes \myone_B \ \ket{\Phi_{AB}^+} = \ket{D}_A
\ket{+x}_B$ and therefore $ \widetilde{\rho}_{AB}^{decoy} = 2
\ket{D}\bra{+x} \otimes \myone \ \widetilde{\rho}_{AB}^{bit} \
\ket{+x}\bra{D} \otimes \myone$.

\subsubsection{Characterizing the set $\Gamma_{Q,V}$}

The set $\Gamma_{Q,V}$ contains any state of the form \be
\sigma_{AB} = \lambda_1 P_{\Phi^+} + \lambda_2 P_{\Phi^-} +
\lambda_3 P_{\Psi^+} + \lambda_4 P_{\Psi^-} \ee where we use the
notation $P_{\Phi} = \ket{\Phi}\bra{\Phi}$ for any state
$\ket{\Phi}$, where the $\ket{\Phi^\pm}, \ket{\Psi^\pm}$ are the
Bell states, and where \ba
\begin{array}{rclrcl} \lambda_1 & = & \bra{\Phi^+} \ \rho \
\ket{\Phi^+}, \quad & \lambda_2 & = & \bra{\Phi^-} \ \rho \
\ket{\Phi^-} \\
\lambda_3 & = & \bra{\Psi^+} \ \rho \ \ket{\Psi^+}, \quad &
\lambda_4 & = & \bra{\Psi^-} \ \rho \ \ket{\Psi^-}
\end{array} \label{def_lambdas}.\ea

The first constraint is the definition of the QBER, the same for
all protocols, namely \be Q = \lambda_3 + \lambda_4.
\label{constr_Q}\ee The constraint that defines $V$ is typical of
this protocol. To derive it, we use the fact that the probability
for decoy sequences to be detected correctly by Bob is
$\frac{1+V}{2}$: \ba \frac{1 \pm V}{2} & = & \bra{D} \otimes
\bra{\pm x} \ \widetilde{\rho}_{AB}^{decoy} \ \ket{D} \otimes
\ket{\pm x} \nonumber\\
& = & 2 \ \sandwich{+x, \pm x}{\rho}{+x, \pm x}\,. \ea Since
$\ket{+x,+x} = \frac{1}{\sqrt{2}} (\ket{\Phi^+} + \ket{\Psi^+})$,
then \ba \frac{1+V}{2} & = & (\bra{\Phi^+} +
\bra{\Psi^+}) \rho (\ket{\Phi^+} + \ket{\Psi^+}) \nonumber\\
& = & \lambda_1 + \lambda_3 + \left(\bra{\Phi^+} \ \rho \
\ket{\Psi^+}+c.c.\right)\,.\ea The Cauchy-Schwartz inequality
implies $|\bra{\Phi^+} \ \rho \ \ket{\Psi^+}| \leq \sqrt{\lambda_1
\lambda_3}$, and therefore $|\bra{\Phi^+} \ \rho \ \ket{\Psi^+} +
\bra{\Psi^+} \ \rho \ \ket{\Phi^+}| \leq 2 \sqrt{\lambda_1
\lambda_3}$. We finally obtain the following constraint: \be
(\sqrt{\lambda_1} - \sqrt{\lambda_3})^2 \leq \frac{1+V}{2} \leq
(\sqrt{\lambda_1} + \sqrt{\lambda_3})^2\,. \label{constr_V1}\ee
Similarly, starting from $\frac{1-V}{2}$, one obtains \be
(\sqrt{\lambda_2} - \sqrt{\lambda_4})^2 \leq \frac{1-V}{2} \leq
(\sqrt{\lambda_2} + \sqrt{\lambda_4})^2\,. \label{constr_V2}\ee

For a state $\sigma_{AB}$ to be in the set $\Gamma_{Q,V}$, its
coefficients $\lambda_s$ therefore have to satisfy the constraints
(\ref{constr_Q}), (\ref{constr_V1}) and (\ref{constr_V2}), along
with the normalization condition $\lambda_1 + \lambda_2 +
\lambda_3 + \lambda_4 = 1$.

\subsubsection{Lower bound on the secret key rate}

\begin{center}
\begin{figure}
\includegraphics[width=8cm]{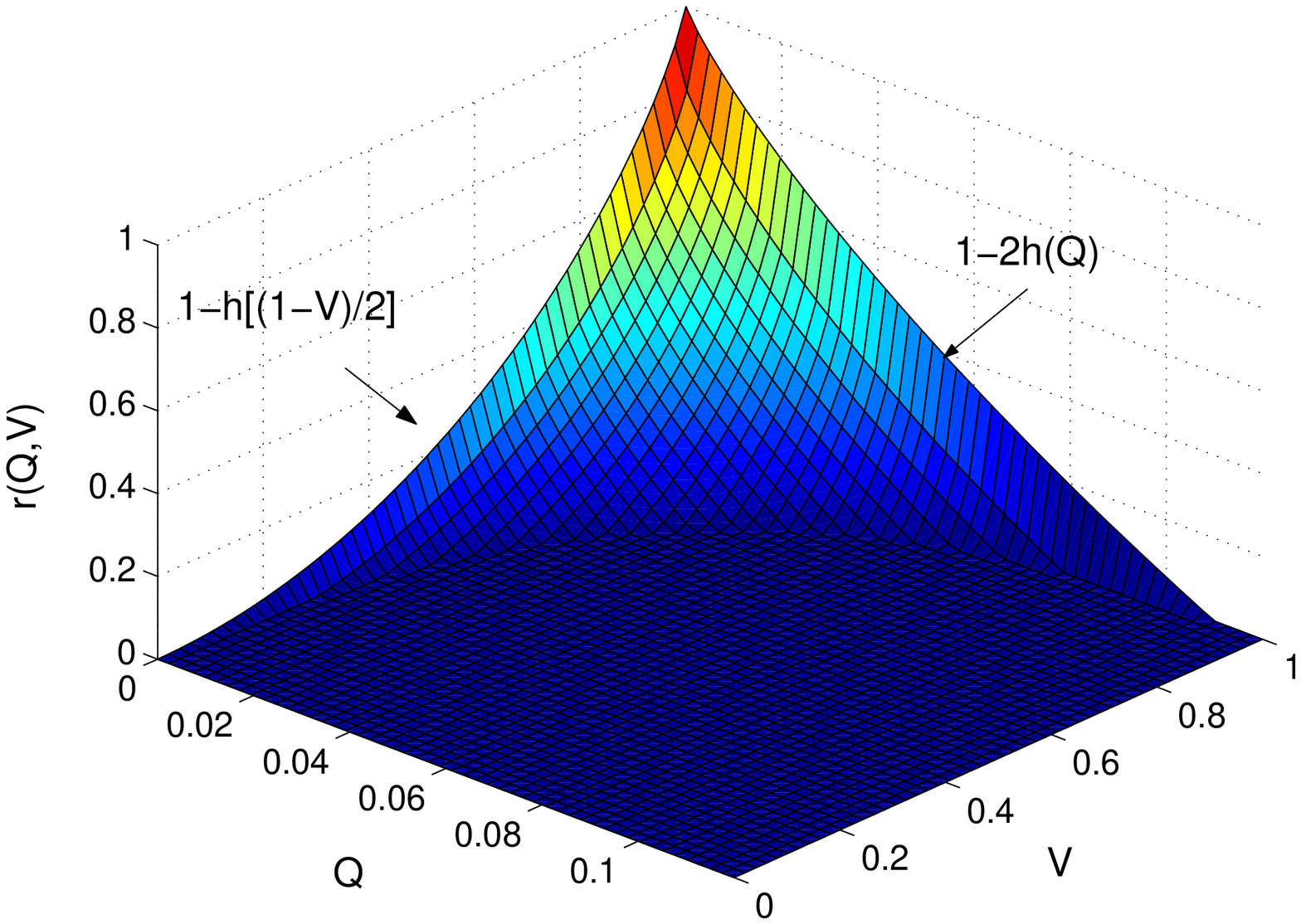} \vspace{2mm}
\includegraphics[width=7cm]{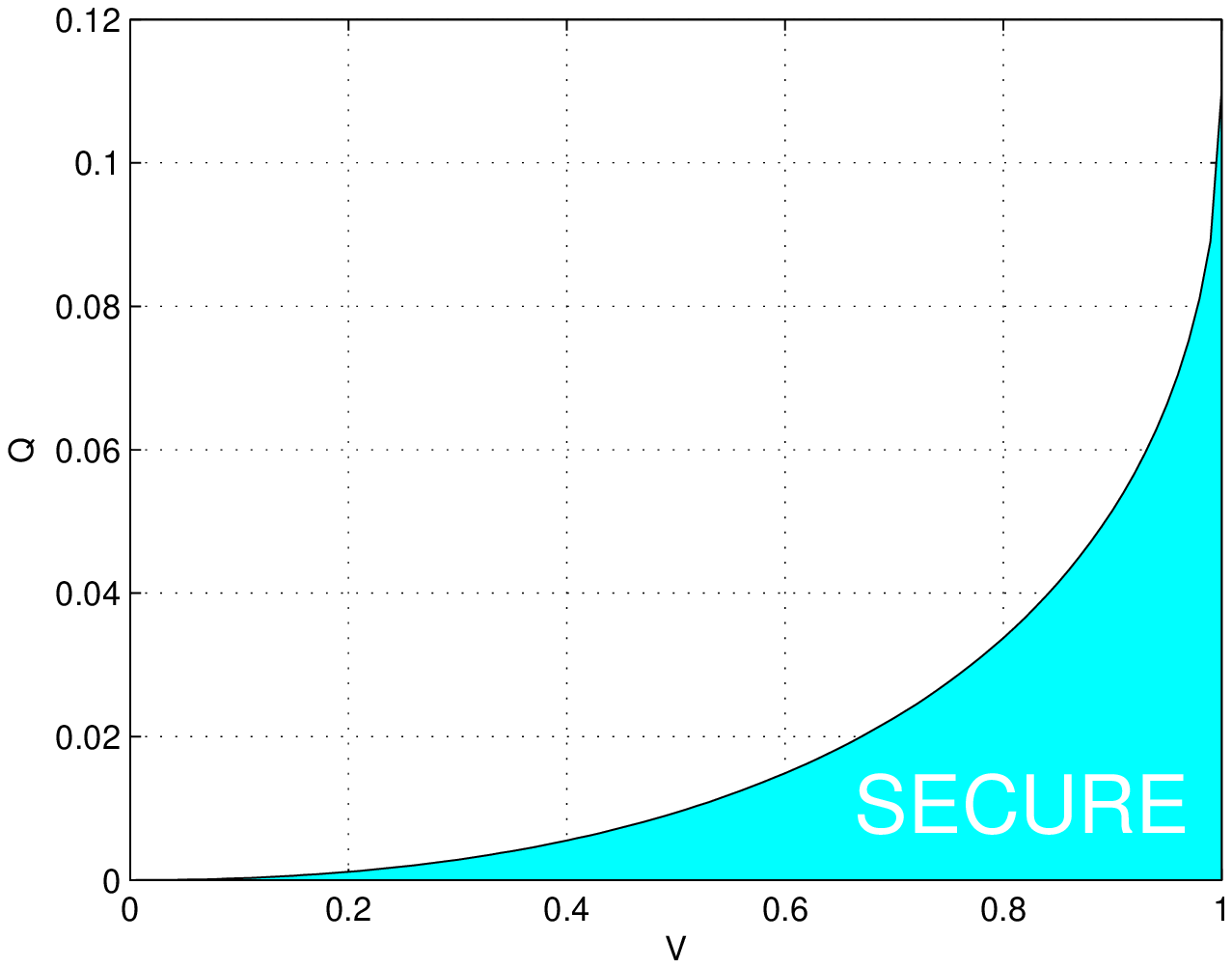}
\caption{Security study of the three-state protocol in a
single-photon implementation. Upper graph: lower bound $r$ as a
function of $Q$ and $V$; lower graph: projection of the upper
graph on the $(Q,V)$ plane, showing the region of parameters in
which the protocol is provably secure.} \label{fig_R_3D}
\end{figure}
\end{center}

Now we have to compute the bound (\ref{lower_bound}). One can show
that, given our constraints, the infimum of $1-S(\si_{AB})$ is
obtained when \ba \sqrt{\lambda_1} + \sqrt{\lambda_3} & = &
\sqrt{\frac{1+V}{2}}
\\ \sqrt{\lambda_2} - \sqrt{\lambda_4} & = & \sqrt{\frac{1-V}{2}}
\ea These equalities, together with Eq.~(\ref{constr_Q}) and the
normalization condition, allow an analytical expression of the
lower bound: \ba r(Q,V)&\geq& 1-H\left(\big[\lambda_1,
\lambda_2,\lambda_3,\lambda_4\big]\right) \label{rexpl}\ea with
\ban
\begin{array}{lcl} \lambda_1 &=& (1-Q) \left[ \frac{1+V}{2} - QV -
\sqrt{(1-V^2)Q(1-Q)} \right]\,, \\
\lambda_2 &=& (1-Q) \left[ \frac{1-V}{2} + QV +
\sqrt{(1-V^2)Q(1-Q)} \right]\,, \\
\lambda_3 &=& Q \left[ \frac{1-V}{2} + QV +
\sqrt{(1-V^2)Q(1-Q)} \right]\,, \\
\lambda_4 &=& Q \left[ \frac{1+V}{2} - QV - \sqrt{(1-V^2)Q(1-Q)}
\right]\,. \end{array}\ean The results are plotted in
Fig.~\ref{fig_R_3D}. For all values of the parameters, the rates
we find are equal or better than those found by Fung and Lo
\cite{funglo}: in particular, for $V=1$ we find security up to
$Q\approx 11\%$, while they reach only up to $Q\lesssim 7.57\%$
(see Fig.~2 of Ref.~\cite{funglo}, where
$\alpha\equiv\frac{1-V}{2}$ and $e_b\equiv Q$).

\subsubsection{Special cases $Q = 0$, $V = 1$}

Let's study the particular cases $Q = 0$ and $V = 1$. With the
previous analysis, we find \ba R(Q=0,V) &\geq& 1 -
h\left(\frac{1-V}{2}\right)\,,\\ R(Q,V=1)& \geq& 1 - 2 h(Q) \ea
where $h$ is binary entropy. In particular, the second rate is the
same as the one obtained for the BB84 protocol \cite{sp}.

In these limiting cases, we have been able to compute a lower
bound in a different way, namely using the Devetak-Winter bound
for collective attacks \cite{dw} and then invoking a de Finetti
theorem to extend the result to all possible attacks
\cite{rennerthesis}. For the case $Q=0$, we find exactly the same
result; for the case $V=1$ however, the lower bound calculated in
this new way is slightly better. This is not a contradiction, as
the method of Ref.~\cite{krausrenner} is not claimed to provide
tight bounds in all circumstances.

\subsection{Weak Coherent Pulses}

\subsubsection{Conservative lower bound}

In our three-state protocol, exactly as it happens for BB84, as
soon as a pulse contains two photons, Eve can obtain full
information using the PNS attack. Therefore, all the pulses
containing more than one photon are "tagged": it is as if they
would carry a label which reveals the quantum state. Once one has
a lower bound $r$ in a single-photon implementation, a lower bound
for implementations with weak coherent pulses can be computed
using the techniques developed in Ref.~\cite{gllp}.

Let $\Delta$ be the fraction of tagged signals: on these, Eve has
full information thanks to the tag. Eve's best strategy consists
in introducing no error on the tagged pulses, and a larger error
$Q_1=\frac{Q}{1-\Delta}$ on the untagged ones, so that the total
QBER is still $Q$. A similar reasoning holds for $V$: in Eve's
best strategy, the tagged pulses have $V=1$, therefore the single
photon pulses have $V_1=\frac{V-\Delta}{1-\Delta}$. These
estimates have a bearing on privacy amplification, while error
correction must be done for the average $Q$. The achievable secret
key rate is finally bounded as \ba r\geq
\,\big[(1-\Delta)\,S\left(Q_1,V_1\right)-h(Q)\big]
\label{boundgllp}\ea where $S(Q,V)=r(Q,V)-h(Q)$ and $r(Q,V)$ is
the single-photon lower bound of Eq.~(\ref{rexpl}). Finally, it is
easy to compute the optimum value of $\Delta$. In general,
$\Delta$ is the probability that Alice sends more than one photon,
conditioned to the fact that Bob has received something. Clearly,
the best case for Eve is that Bob {\em always} receives something
when Alice has sent two or more photons. Therefore \ba
\Delta&=&\frac{1-e^{-\mu}-\mu e^{-\mu}}{1-e^{-\mu
t\eta}}\,\approx\,\frac{\mu}{2t\eta}\,. \label{delta}\ea Knowing
this, one can now multiply $r$ by Bob's detection rate to obtain
the secret key rate in bits per pair of pulses, then optimize
$\mu$ to maximize this quantity. Note that the lower bound
(\ref{boundgllp}) is very conservative because it holds only for
the untrusted-device scenario --- this is why the denominator in
(\ref{delta}) contains $\eta$ as well; it is not known how to
prove a rigorous lower bound in the trusted-device scenario. (See
also \cite{finitedetectors}).

\subsubsection{PNS attack in the zero-error case}
\label{sspns}

In the main text, we have presented zero-error attacks against the
COW protocol in the trusted-device scenario. For comparison, we
compute the PNS attack against the three-state protocol
implemented with weak coherent pulses: we recall that in this
protocol, contrary to COW, there is no coherence across the bit
separation.

If $Q=0$ and $V=1$, we have $I(A:B)=1$. Eve counts the number of
photons in each two-pulse sequence corresponding to a bit: if she
finds $n=1$, she can either let the photon go or block it, but in
any case she can't learn anything; if she finds $n>1$, she keeps
some photons and sends the others to Bob, and she has full
information. For the purpose of this simple analysis, we write
everything in the case $\mu\ll 1$ (the generalization is
straightforward but complicates the formulae). We have then
$I(A:E)=\frac{\mu}{2t}$, the difference with (\ref{delta}) coming
from the fact that we can compute this upper bound in the
trusted-device scenario. The rate per bit becomes \ba R&= &
\left(1-\frac{\mu}{2t}\right)\,\mu\, t\,t_B\eta(1-f)\,.\ea This
expression is optimal for $\mu_{opt}=t$, therefore \ba
R(\mu_{opt})&\approx&\frac{t^2}{2}\,t_B\eta(1-f)\,. \ea This
scales as $t^2$, as it happens for BB84 under the same conditions
\cite{armand}. This rate is much smaller than the upper bounds
obtained in the main text for the COW protocol for the most
powerful attacks described in this paper (Fig.~\ref{figmain}). A
better attack may exist against COW; however, we conjecture that
this difference is intrinsic
--- in physical terms, we conjecture that the existence of
coherence across the bit separation is a real advantage and
increases the extractable secret key rates by a significant
amount.

\section{Beam-Splitting attack and Devetak-Winter bound}
\label{App_BS_Holevo}

The beam-splitting attack is the only known attack which will
simulate exactly all statistics for Alice and Bob given a lossy
channel, since it is a physical model for such a lossy channel.
The fraction $1 - t$ of lost photons are given to Eve, who has
forwarded the remaining fraction $t$ to Bob through a lossless
channel. The information that Eve can extract from her data
depends on the way she processes them. For each bit she wants to
estimate, Eve faces the problem of distinguishing the two states
$\ket{0 \alpha'}$ and $\ket{\alpha' 0}$, where $\alpha' =
\sqrt{1-t}\, \alpha$.

In Refs \cite{COW_1,COW_2}, it was supposed that Eve performed the
same measurement as Bob: she measures the time of arrival for each
pulse, which corresponds to the best unambiguous state
discrimination between the two states $\ket{0 \alpha'}$ and
$\ket{\alpha' 0}$. With probability $1 - \braket{0
\alpha'}{\alpha' 0}$, the result is conclusive and she gets full
information on the bit. Her average information on each bit is
then \ba I_{USD} &=& 1 - \braket{0 \alpha'}{\alpha' 0}\,. \ea
However, there are other measurements that could give Eve more
information. For instance, the minimum-error measurement
\cite{helstrom} would give her the information \ba I_{ME} & =& 1 -
h \left( \frac{1}{2} - \frac{1}{2} \sqrt{1 - \braket{0
\alpha'}{\alpha' 0}^2} \right)\,,\ea which is larger than
$I_{USD}$ ($h$ is the binary entropy function).

The USD and ME measurements are bitwise measurements, and define
the so-called individual (or incoherent) attacks. More generally,
Eve can be allowed to make a {\em collective} attack from
beam-splitting: each signal is split with the same fraction, as
above, but then Eve is allowed to wait until the end of classical
post-processing (error correction, privacy amplification) before
performing a (possibly coherent) measurement on the quantum
systems she has kept. What Eve does maybe hard to find (actually,
to our knowledge, this is not known for any protocol); but a
computable bound for the secret key rate that can be extracted in
the presence of collective attacks has nevertheless be found by
Devetak and Winter \cite{dw}. The upper bound on the accessible
information that Eve can learn, whatever the measurement she
performs, is given by the Holevo bound \cite{Holevo}. For the
problem of distinguishing the two states $\ket{0 \alpha'}$ and
$\ket{\alpha' 0}$, the Holevo bound is \cite{noteholevo} \ba
\chi_{Hol} &=& h \left(\frac{1 - \braket{0 \alpha'}{\alpha' 0}}{2}
\right)\,.\ea The Devetak-Winter bound for the secret key rate
reads then \ba R &\geq &(1-f) \,\left(1 - e^{-\mu t t_B
\eta}\right)\, \left(1 - \chi_{Hol}\right)\\ & \gtrsim & (1-f)
\,\mu t t_B \eta\, \left[1 - h \Big(\frac{1 - e^{-\mu (1-t)}}{2}
\Big)\right]\ea the second expression being for the case $\mu t
t_B \eta \ll 1$.

As usual, Alice and Bob should choose $\mu$ in order to optimize
$R$. Let's define $g(x) = x \big[1 - h(\frac{1 -
e^{-x}}{2})\big]$. Numerically, we find $\sup_x g(x)\equiv g(\xi)
\approx 0.1428$, obtained for $\xi \approx 0.4583$. Therefore, the
optimal value of $\mu$ in the case of a collective beam-splitting
attack is \ba \mu_{opt} &=& \frac{\xi}{1-t}\ea and the
corresponding lower bound on the extractable secret key rate is
\ba R(\mu_{opt})& =& g(\xi)\, \frac{t}{1-t} \, t_B \eta (1-f)\,.
\ea This is what we plotted in Figs \ref{figmain} and
\ref{figs_att3} in comparison to our attacks.

\section{On the attack that reproduces the detection rates}
\label{appqs}

We give here the calculation of $(q_0,q_1,q_2,q_3)$ that define
the attack that reproduces the detection rates studied in Section
\ref{secrates}, and comment on some of its features. We recall
that we work in the limit $\mu t \eta\ll 1$ and that we suppose
that Eve sends one photon to Bob when she has got a conclusive
result.

\subsection{Calculation of the parameters $(q_0,q_1,q_2,q_3)$ of the attack}

For $D_{B,bit}$ and $D_{B,decoy}$, the requirement
(\ref{requirement}) leads respectively to the following two
conditions: \ba \mu(t-q_0)&=&
\frac{\frac{2}{3}\,q_1\,p_{concl}^{0\alpha
0}+\frac{1}{2}\,q_2\,p_{concl}^{0\alpha:\alpha 0}}{1-f}
\,,\label{cond1}\\
\mu(t-q_0)&=& \frac{q_3\,p_{concl}^{0:\alpha\alpha:
0}}{4f}\,.\label{cond2}\ea Given these two conditions, the
requirement (\ref{requirement}) is automatically satisfied for
$D_{Mj,2k}$ for both $j=1,2$. This is not astonishing, as these
detection rates depend on $f$ in the same way as those of $D_B$
do. Finally, for the $D_{Mj,2k+1}$, the requirement
(\ref{requirement}) gives two new conditions: \ban
\mu(t-q_0)&=&\frac{ \frac{4}{3}\,q_1\,p_{concl}^{0\alpha
0}+2q_2\,p_{concl}^{0\alpha:\alpha 0}+
q_3\,p_{concl}^{0:\alpha\alpha: 0}}{(1+f)(3+f)}\,,\\
\mu(t-q_0)&=&\frac{ \frac{4}{3}\,q_1\,p_{concl}^{0\alpha 0}+
q_3\,p_{concl}^{0:\alpha\alpha: 0}}{1-f^2}\,. \ean It can be
checked that one of these conditions is redundant, as it follows
exactly from assuming the other one together with (\ref{cond1})
and (\ref{cond2}); as a third condition, we take then a simple
linear combination of the last two ones, which reads \ba
\mu(t-q_0)&=& \frac{q_2\,p_{concl}^{0\alpha:\alpha 0}}{(1+f)^2}\,.
\label{cond3}\ea In summary, we have four linear conditions
[(\ref{cond1}), (\ref{cond2}), (\ref{cond3}) and the normalization
(\ref{norm})] for the four coefficients $q_j$: the system can be
solved exactly as a function of $\mu$, $t$ and $f$.

The solution --- whose result is given in the main text, Eqs
(\ref{solqs})--(\ref{f3}) --- goes as follows. For $j=1,2,3$, we
have $q_j=\mu(t-q_0)F_j$ where $F_2$ can be read directly in
Eq.~(\ref{cond3}), $F_3$ in Eq.~(\ref{cond2}), and
$F_1=3(1-4f-f^2)/4p_{concl}^{0\alpha 0}$ can be derived from those
and from Eq.~(\ref{cond1}). The normalization condition
(\ref{norm}) gives then $q_0 = \frac{\mu tF-1}{\mu F-1}$ with
$F=F_1+F_2+F_3$.

We must still verify that $q_0$ is a probability. Since $t<1$, the
condition $q_0\leq 1$ is satisfied provided $\mu F>1$, which is
true for all values of $\mu$ and $f$ (in fact, it can be verified
that the minimal value of $\mu F$, obtained for $\mu\approx 2$, is
of the order 100, slightly dependent on $f$). Given $\mu F>1$, the
condition $q_0\geq 0$ is satisfied provided $\mu t F\geq 1$. To
fulfill this condition, one must know how $\mu$ varies with $t$.
Let's consider first $\mu_{opt}$ as defined in (\ref{muopt}): then
$\mu t F=3(1-t)$, therefore the condition is satisfied for $t\leq
\frac{2}{3}$ or (with the parameters used for the graphs)
$\ell\gtrsim 7$km --- in practice, recall that (\ref{muopt}) is
valid for $\mu\ll 1$ that is for $t\ll 1$; so the result is
consistent. If we take now $\mu_{max}=\sqrt{3} \mu_{opt}$, we find
$\mu t F=1-t$: the condition can never be satisfied. This is not
really a problem: it simply means that Eve must add some losses,
i.e. that we must add to her strategy the possibility of blocking
pulses.

\subsection{Behavior of $q_1,q_2,q_3$}

In general, it holds $F_3>F_2>F_1$, that is, $q_3>q_2>q_1$, for
all values of $f$ and $\mu$. The fact that $q_3$ does not vanish
(and remains even larger than $q_1$ and $q_2$) if $f\equiv 0$ is
an artefact of the solution of the system. In fact, the
requirement on $D_{B,decoy}$ reads originally $4f\,\mu(t-q_0)\,=\,
q_3\,p_{concl}^{0:\alpha\alpha: 0}$: if $f>0$, it gives
(\ref{cond2}) as we stated it; but if $f=0$, the requirement is
automatically satisfied and no constraint is put on $q_3$ (the
best choice for Eve would then be $q_3=0$). In any case, COW
without decoy sequences would be much more vulnerable against
Eve's attacks \cite{COW_1,COW_2}, so the case $f\equiv 0$ is not
of real interest. A more meaningful question is, what happens in
the limit $f\rightarrow 0$ for {\em real} implementations (blurred
statistics, finite key length); but, as already mentioned, we
haven't developed the mathematical tools yet, which would allow to
tackle this problem.

\section{The consequence of poor statistical analysis: an example}
\label{appexample}

Let us suppose that Alice and Bob verify $Q=0$, $V=1$ (without
distinguishing decoy sequences from $1-0$ bit sequences) and just
the average detection rate $D_B^t$. In particular, they don't
check that the fraction of decoy sequences is the expected one:
Eve can set $q_3=0$. As simple examples of the attacks that become
possible, Eve can always attack with USD3 ($q_2=0$) or with USD4a
($q_1=0$).

\begin{center}
\begin{figure}
\includegraphics[width=8cm]{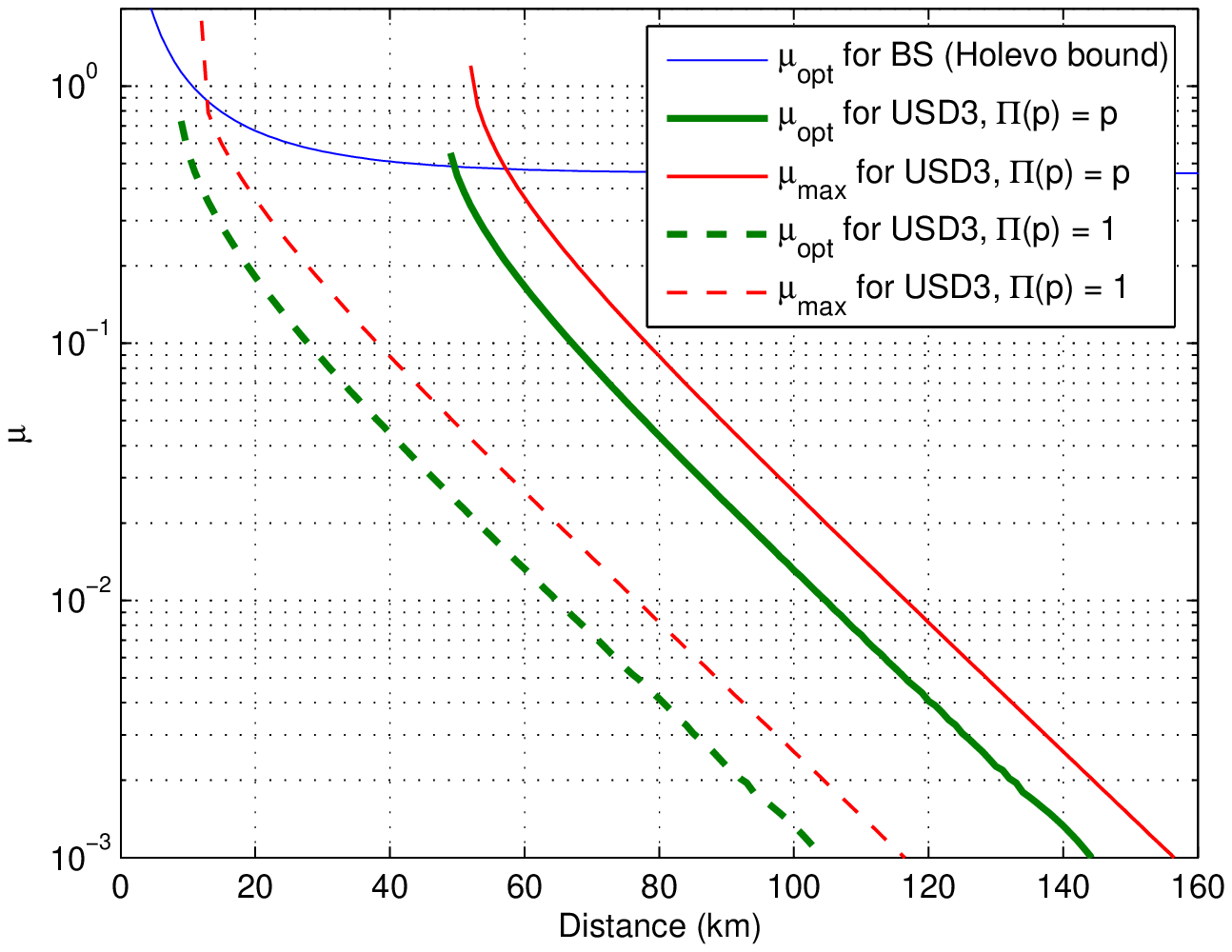} \vspace{2mm}
\includegraphics[width=8cm]{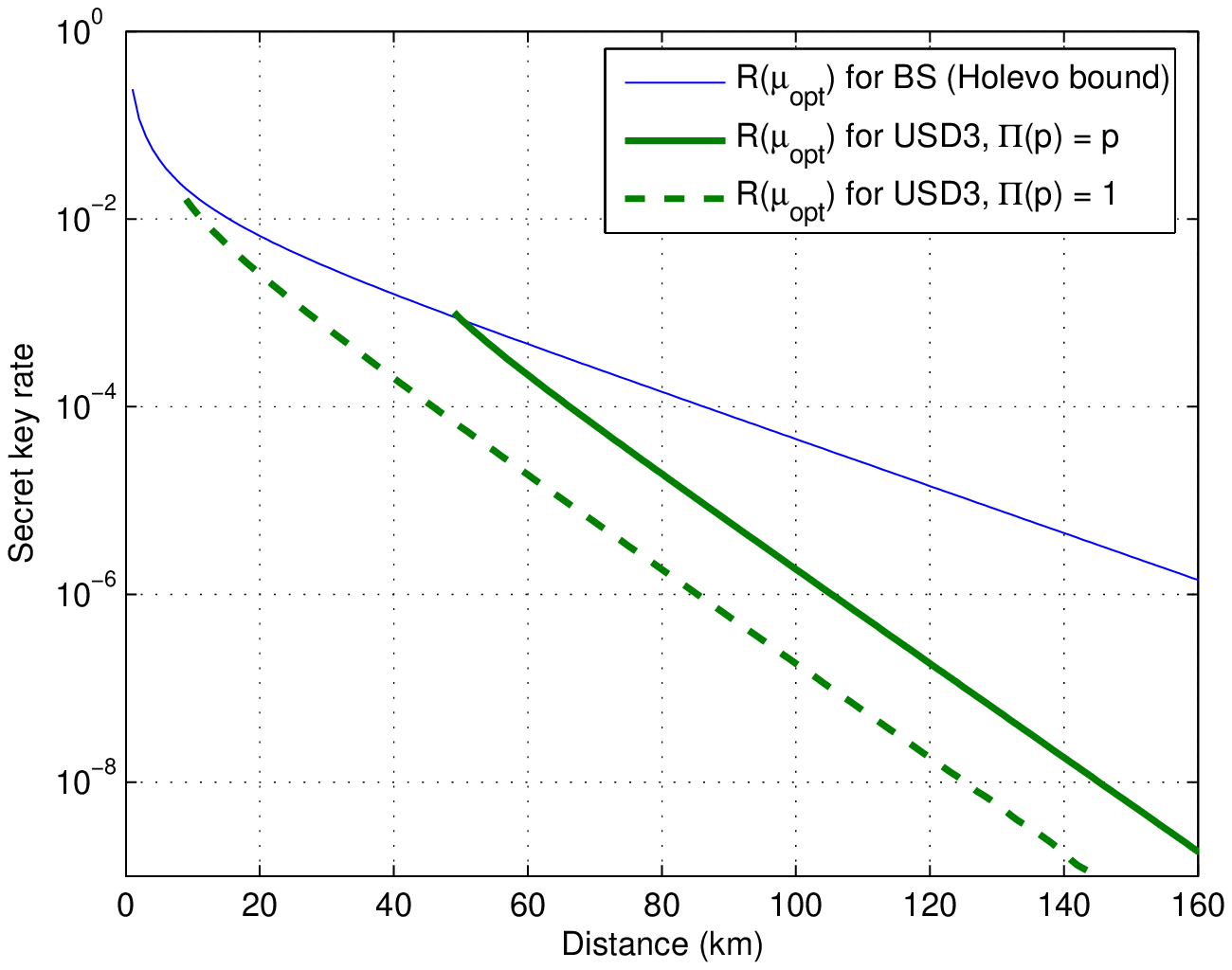}
\caption{USD3 attack, which becomes possible if Alice and Bob
check only the average detection rate. We plot the optimal mean
photon number $\mu_{opt}$ (upper graph) and corresponding secret
key rate $R$ (lower graph) as a function of the distance $d$. Full
lines: results for $\Pi(t_B\eta)=t_B\eta$ (Eve forwards one
photon); dashed lines: results for $\Pi(t_B\eta)=1$ (Eve forwards
bright pulses). The attack is again compared to the Holevo bound
on the BS attack (Appendix \ref{App_BS_Holevo}). The parameters
are the same as in Fig.~\ref{figmain}.} \label{figs_att3}
\end{figure}
\end{center}

{\em USD3 attack.} If $q_2=q_3=0$ and only the detection rate in
$D_B$ is monitored, the set of requirements (\ref{requirement})
reduce to the sole condition $q_1 D_B^{(3)} + (1-q_1) D_B^{t = 1} =
D_B^t$ i.e. \ba q_1& =& \frac{D_B^{t=1} - D_B^t}{D_B^{t=1} -
D_B^{(3)}}\,. \ea The secret key rate that can be extracted against
such an attack is \ba R &=&
(1-q_1)D_{B,bit}^{t=1}\,=\,\left(\frac{D_B^{t} -
D_B^{(3)}}{D_B^{t=1} - D_B^{(3)}}\right)\,D_{B,bit}^{t=1}\,. \ea The
values of $\mu_{max}$, $\mu_{opt}$ and $R(\mu_{opt})$ can now be
computed as a function of $t$. Numerical solutions are plotted in
Fig.~\ref{figs_att3}, as a function of the distance. We have plotted
two series of curves for our attack (describing the cases where Eve
forwards either one photon or bright pulses) against the curve
associated to the BS attack. Analytical solutions can be obtained in
the limit $\mu<<1$: $\mu_{max} = Ct$, $\mu_{opt}=Ct/2$ and
$R(\mu_{opt})=\frac{1-f}{4} t_B \eta \ C t^2$ with
$C=[6(1+f)t_B\eta]/[(1-f)^2\,\Pi(t_B \eta)]$. Note that
$R(\mu_{opt})\propto t^2$, whereas for the attack that preserves the
detection rates we had the much slower decrease $R(\mu_{opt})\propto
t^{3/2}$ [Eq.~(\ref{Rmax})].

{\em USD4a attack.} The analysis of the case $q_1=q_3=0$ follows
exactly the same pattern, just replacing $D_B^{(3)}$ with
$D_B^{(4)}$ --- in fact, the only difference is the factor
$\frac{4}{3}$ which relates these two quantities, see Eqs
(\ref{db3}) and (\ref{db4a}). This attacks gives slightly better
rates than those plotted in Fig.~\ref{figs_att3}; in the case
$\mu<<1$, the analytical solutions for $\mu_{max}$, $\mu_{opt}$
and $R(\mu_{opt})$ are the same as before, with now
$C=[8(1+f)t_B\eta]/[(1-f)^2\Pi(t_B\eta)]$.

The message of Fig.~\ref{figs_att3} is clear: these attacks are
significantly more powerful than the one in which Eve is asked to
reproduce all the detection rates (Fig.~\ref{figmain}). In
particular, the distance $\ell$, at which the attacks become
important, is approximately 50km, well within the actual
experimental working range. To avoid these attacks, it is
therefore mandatory that Bob checks carefully his detection rates.

\section{USD attacks in the case of "empty decoy sequences"}
\label{App_empty_decoy}

In this Appendix, we study a modification of the COW protocol,
which makes it more robust against the attacks known to date (in
particular, against the attacks studied in this paper), while
keeping the simplicity at the experimental level. The modification
consists in introducing a new type of decoy sequence, which is
just {\em two empty pulses}. In this modified COW, Alice sends an
"empty decoy sequence" $\ket{00}$ with probability $f_0$, and a
"full decoy sequence" $\ket{\alpha\alpha}$ with probability $f_1$.
We will write $f = f_0 + f_1$. With probability $\frac{1-f}{2}$,
Alice sends a logical bit 0 (resp. 1).

It may be at first sight astonishing, that additional vacuum
signals may provide an advantage; still, this happens also in
decoy state protocols \cite{decoy}. In our case, the possibility
of new signals (albeit empty ones) makes the unambiguous state
discrimination that we have studied in Section \ref{sec3} less
efficient, because the set of possible states becomes larger.

\subsection{Attack on 3 pulses}

Eve wants to discriminate the state $\ket{0 \alpha 0}$ from the {\em
seven} other possible states, which are now: \be \ket{0 0 0}\,,\;
\ket{0 0 \alpha}\,,\; \ket{0 \alpha \alpha}\,,\; \ket{\alpha 0
0}\,,\; \ket{\alpha 0 \alpha}\,,\; \ket{\alpha \alpha 0}\,,\;
\ket{\alpha \alpha \alpha}\,. \label{7states}\ee Note that the
previous state $\ket{\psi_{0 \alpha 0}}$ [Eq.~(\ref{psi3})] is not
orthogonal to $\ket{0 0 0}$. Instead, the state orthogonal to the
seven states listed in (\ref{7states}) is \ba \ket{\psi_{0 \alpha
0}} &=& \frac{\phi(\alpha) - \chi \phi(0)}{\sqrt{1-\chi^2}} \ea
where $\phi$ is given by Eq.~(\ref{def_phi}). As before, Eve
performs a projective measurement which separates $\ket{\psi_{0
\alpha 0}}$ from the subspace orthogonal to it. Conditioned on the
fact that the state $\ket{0 \alpha 0}$ was sent by Alice, the
probability of a conclusive result is $\left|\braket{0 \alpha
0}{\psi_{0 \alpha 0}}\right|^2 = (1 - \chi^2)^3=(1 - e^{-\mu})^3$.
This is smaller than the value $(1 - e^{-\mu})^2$ found in the
absence of empty decoy sequences.

\subsection{Attack on 4 pulses}

Eve wants to discriminate the state $\ket{0 \alpha \alpha 0}$ from
the {\em fifteen} other possible states, which are now: \be
\begin{array}{c} \ket{0 0 0 0}, \ket{0 0 0 \alpha}, \ket{0 0 \alpha 0}, \ket{0 0 \alpha \alpha}, \ket{0 \alpha 0 0},
\\ \ket{0 \alpha 0 \alpha}, \ket{0 \alpha \alpha \alpha}, \ket{\alpha 0 0 0}, \ket{\alpha 0 0 \alpha}, \ket{\alpha 0 \alpha 0},
\\ \ket{\alpha 0 \alpha \alpha}, \ket{\alpha \alpha 0 0}, \ket{\alpha \alpha 0 \alpha},
\ket{\alpha \alpha \alpha 0}, \ket{\alpha \alpha \alpha \alpha}.
\label{15states} \end{array} \ee Note that the analysis is the
same for attacks USD4a and USD4b here, since all the sequences are
possible.

The state orthogonal to these fifteen states is \ba \ket{\psi_{0
\alpha \alpha 0}} & = &\frac{\phi(\alpha\alpha)-\chi \phi(0
\alpha)-\chi \phi(\alpha 0) + \chi^2 \phi(\alpha
\alpha)}{1-\chi^2}. \ea Conditioned on the fact that the state
$\ket{0 \alpha \alpha 0}$ was sent by Alice, the probability of a
conclusive result is $\left|\braket{0 \alpha \alpha 0}{\psi_{0
\alpha \alpha 0}}\right|^2 = (1 - \chi^2)^4$. Again, the
probability of success is smaller than the probability of success
$\frac{(1 - \chi^2)^3}{1 + \chi^2}$ for the USD4b attack, and much
smaller than the one $(1 - \chi^2)^2$ for the USD4a attack in the
absence of empty decoy sequences.

\subsection{Attack that preserves the detection rates}

The study follows exactly the same lines as for the attack studied
in Section \ref{secrates} and Appendix \ref{appqs}. As we did
there, we suppose that Eve performs one of the three USD attacks
with probabilities $q_j$, or forwards the pulses through a
lossless channel with probability $q_0$. The probabilities for
each USD attack to be conclusive are the following : \ba
p_{concl}^{0\alpha0} &=&
\frac{1-f}{2}\left(\frac{1-f}{2}+f_0\right) (1
- e^{-\mu})^3\,, \\
p_{concl}^{0\alpha:\alpha0} &=& \left(\frac{1-f}{2}\right)^2 (1 -
e^{-\mu})^4\,,
\\p_{concl}^{0:\alpha\alpha:0} &=& f_1 \left(\frac{1-f}{2}+f_0\right)^2 (1 -
e^{-\mu})^4\,. \ea Under the assumption that Eve forwards one
photon when her attack is conclusive, and in the regime where $\mu
\eta \ll 1$, one finds $q_j = \mu (t-q_0) F_j$ for $j = 1,2,3$,
and $q_0 = \frac{\mu t F - 1}{\mu F - 1}$, with now:
\ba F_1 &=& \frac{3(1 - 4f_1 - (f_1 - f_0)^2)}{4p_{concl}^{0\alpha0}} \\
F_2 &=& \frac{(1 - f_0 + f_1)^2}{p_{concl}^{0\alpha:\alpha0}} \\
F_3 &=& \frac{4f_1}{p_{concl}^{0:\alpha\alpha:0}} \\
F &=& F_1 + F_2 + F_3\,. \ea Apart from the obvious restriction
$f_0+f_1\leq 1$, since $F_1$ has to be positive there is a
restriction on the values of $f_0$ and $f_1$ for this attack to be
possible: $f_1 \leq \min\left(1/4, - 2 + f_0 + \sqrt{5 -
4f_0}\right)$.

The upper bound on the extractable secret key rate is \ba R(\mu)& =&
q_0 D_{B,bit}^{t=1} \,=\, q_0 \mu t_B \eta (1-f)\,.\ea In the limit
$\mu \ll 1$, the optimization of $R$ can be done analytically, using
$F(\mu) \approx \frac{4 {\cal F}}{\mu^4}$ with ${\cal F} = \frac{(1
+ f_1 - f_0)^2}{(1 - f_1 - f_0)^2} + \frac{4}{(1 - f_1 + f_0)^2}$
and $q_0 \approx t - \frac{\mu^3}{4{\cal F}}$. In order to optimize
$R$, Alice and Bob will choose \ba \mu_{opt} &\approx& {\cal
F}^{1/3}\, t^{1/3}\ea and obtain the rate \ba R(\mu_{opt}) &\approx&
\frac{3 {\cal F}^{1/3}}{4} t_B \eta (1-f) \, t^{4/3}\,.
\label{rempty}\ea Note that now, $\mu_{opt} \propto t^{1/3}$ and
$R(\mu_{opt}) \propto t^{4/3}$: the new protocol with empty decoy
sequences is more robust against our USD attacks. Besides, one gets
$\mu_{max} = 4^{1/3}\mu_{opt}$.

\begin{center}
\begin{figure}
\includegraphics[width=8cm]{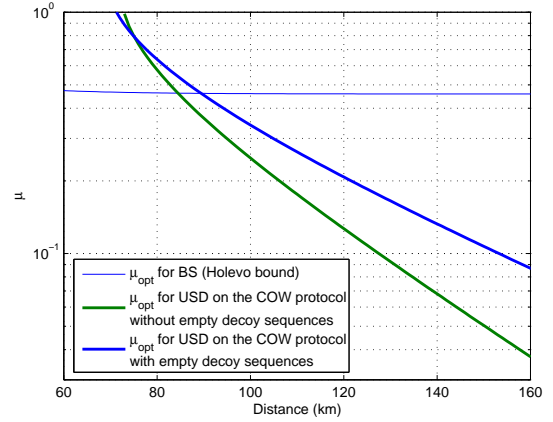} \vspace{2mm}
\includegraphics[width=8cm]{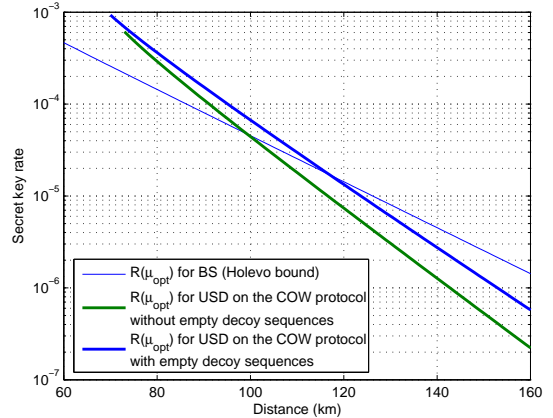}
\caption{USD attack that reproduces the detection rates, on the COW
protocol, with and without empty decoy sequences, compared to the
Holevo bound on the BS attack. Same parameters as in
Fig.~\ref{figmain}, and $f_0=f_1=0.05$.} \label{fig_empty_decoy}
\end{figure}
\end{center}

In general, the optimization of $R$ over $\mu$ must be done
numerically. We show the results in Fig.~\ref{fig_empty_decoy} for
the same parameters as we used for Fig.~\ref{figmain}, but here
$f=0.1$ is split into $f_0 =f_1= 0.05$. We see that, in the
presence of empty decoy sequences, the USD attack that reproduces
all rates overcomes the beam-splitting attack only for
$\ell\gtrsim 120$km.

\end{multicols}


\begin{thebibliography}{99}
\bibitem{noteUSD} Usually, when speaking of USD measurements, the
goal is to discriminate all the $n$ possible states. Therefore the
measurement has $n+1$ outcomes, which either identify the state
unambiguously, or say that the discrimination was inconclusive. Such
are in particular the USD attacks as defined in \cite{dusek}.
\bibitem{notefine} The inconclusive result is a projection onto a
subspace. If in this subspace there are still pulse sequences that
are linearly independent from all the others (as is the case in
particular for the USD attacks under study in this paper), each of
these sequences could then in turn be unambiguously discriminated
with some probability.
\bibitem{notedead} If Bob's detectors have a dead time, the
analysis of USD4a is more subtle, because the first non-empty
pulse can be detected with the probability we have written in the
main text; but the following pulse can be detected only if the
first one has not been detected. This may introduce an asymmetry:
the logical bit 1 may be detected more often than the logical bit
0. In the extreme case of bright pulses, the first non-empty pulse
always triggers the detector, therefore only the logical bit 1 is
detected. In the other extreme case, where Eve sends out a single
photon, there is no asymmetry, because only one detection can take
place.
\bibitem{notecoinc} Of course, the fact that Eve forwards always a
single photon can be verified by the absence of the expected
coincidence counts in two or three of Bob's detectors. But this is
not a serious concern: the probability of coincidence is small, and
we can easily suppose that Eve sends sometimes a brighter pulse when
she has a conclusive result in USD3. If she does so, she can
reproduce the coincidence rates. Alice and Bob could still detect
this attack by checking if the cases of coincidences are equally
distributed among all possible bit and decoy sequences; but it is
pointless to make such a detailed analysis here. Finally note that
it may be advantageous for Eve to send other states that a
single-photon state, for instance vacuum-substracted coherent
states.
\bibitem{noteinterf} In the absence of empty decoy sequences, it
never happens that three consecutive pulses are empty; two
additional interferometers, checking the coherence across two and
three time slots respectively, would then be enough to make USD
attacks impossible.
\bibitem{random_transfo} This is after Alice and Bob applied a random permutation on their qubit
pairs and random bit-flip operations, and assuming they apply
optimal error correction and privacy amplification. See
\cite{krausrenner} for details. Note that we don't consider here
the possible classical "preprocessing" $A' \gets A$.
\bibitem{noteholevo} The Holevo bound is computed here for the
task of bitwise distinguishing the states. One can check that the
Holevo bound remains the same if Eve tried to get information on
longer strings of bits.
\end{thebibliography}
\end{document}